\documentclass{article}

\pdfoutput=1
\usepackage{arxiv}

\usepackage[utf8]{inputenc} 
\usepackage[T1]{fontenc}    
\usepackage[hidelinks]{hyperref}       
\usepackage{url}            
\usepackage{booktabs}       
\usepackage{amsfonts}       
\usepackage{nicefrac}       
\usepackage{microtype}      
\usepackage{lipsum}
\usepackage{mathtools}
\usepackage{graphicx}
\graphicspath{ {./images/} }
\usepackage{natbib}
\usepackage{amsmath}
\usepackage{makecell}
\usepackage{adjustbox}
\usepackage[table]{xcolor}
\usepackage{enumitem}

\usepackage{multirow}
\usepackage[title]{appendix}
\usepackage[flushleft]{threeparttable}

\def\bSig\mathbf{\Sigma}

\usepackage{natbib}
\usepackage{amsmath}
\usepackage{mathtools}
\usepackage{makecell}
\usepackage{adjustbox}
\usepackage[table]{xcolor}
\usepackage{siunitx}
\usepackage[mathscr]{euscript}
\usepackage{amsfonts}
\usepackage{xcolor}
\usepackage{xr}

\title{Transporting summary measures of relative effects from randomised trials to the treated patient population: an application to breast cancer endocrine therapy}

\author{Bonnie E. Shook-Sa$^{1,*}$, 
Paul N. Zivich$^{2}$, Carolyn Taylor$^{1}$, David Dodwell$^{1}$,  
\\ \textbf{Jake Probert$^{1}$, Sarah C Darby$^{1}$, and Paul McGale$^{1}$} \\
\\
{$^{1}$Nuffield Department of Population Health, University of Oxford, Oxford, UK} \\
{$^{2}$Department of Epidemiology, University of North Carolina at Chapel Hill,
Chapel Hill, NC, USA} \\
\small{$*$}\footnotesize{Bonnie.Shook-Sa@ndph.ox.ac.uk}}

\begin{document}
 \maketitle
 \begin{abstract}
Randomised trials often report relative treatment effects, such as risk ratios and hazard ratios, for trial populations. Clinical decision-making, however, often benefits from estimates of absolute treatment effects in the population eligible for treatment. Trial participants may not represent this target population well, and restrictions on access to individual participant trial data can further complicate absolute effect estimation. Routine care data are often representative of the target population but may be subject to uncontrolled confounding. We consider estimation of the average treatment effect on the treated (ATT), an absolute measure, by combining a representative sample of treated routine care patients with summary measures (i.e., estimated risk or hazard ratios) from either a randomised trial or a meta-analysis of trials. Under marginal or conditional transportability assumptions, the ATT is shown to be identifiable. The implications of collapsibility of the effect measure on transportability are discussed, and plug-in estimators of the ATT are presented. Simulation studies are used to assess finite sample performance of the estimators in a range of settings. The proposed methods are applied to estimate the ATT of endocrine therapy on 15-year breast cancer mortality using results from a meta-analysis of randomised trials and England's National Disease Registration Service. 
	\end{abstract}

	
\section{Introduction}
In population health, randomised trials are the gold-standard for estimating the benefits or harms of a treatment. While well-conducted trials allow for unbiased estimates of the average treatment effect within the trial population, they often face challenges with external validity. Comparisons of trial participants with real-world cancer patient populations have found that trial participants tend to be younger and healthier, on average \citep{karanis_research_2016,jones_inequalities_2025}. Observational data (e.g., from cancer registries) may better represent the general patient population but analyses of these data must contend with confounding, making robust estimation of treatment effects challenging \citep{breskin_using_2019}.

Several meta-epidemiological and methodological studies have found that relative effects tend to be more stable across participants with differing baseline risks compared to absolute effects \citep{andersen_absolute_2021,schmid_empirical_1998,furukawa_can_2002,Kent_IJE}. When treatment effect heterogeneity on the relative scale is assumed to be absent or restricted to a few measured covariates, researchers may aim to transport relative effect measures across populations. However, even when relative effects are transportable, absolute benefits and harms of treatment will vary depending on individual risk factors. 

In the setting of cancer treatments, quantifying absolute benefits and harms is of key importance for both clinicians and patients. For example, consider estimation of the effect of endocrine therapy on breast cancer mortality among women with oestrogen-receptor positive early breast cancer. Hazard ratios have been estimated in randomised trials and subsequent meta-analyses \citep{early_breast_cancer_trialists_collaborative_group_ebctcg_relevance_2011}. However, absolute benefits of endocrine therapy vary by patient and tumour characteristics. For example, women with positive axillary lymph nodes generally derive greater absolute benefit from a given breast cancer treatment than do women without nodal involvement because their baseline breast cancer mortality risk is higher. Endocrine therapy offers a substantial reduction in relative risk of breast cancer mortality, but there are potential side effects associated with treatment \citep{nci_hormone_therapy_breast_cancer}. When the absolute benefit is small, the balance of benefits and harms may not favour endocrine therapy. Quantifying the absolute benefits of endocrine therapy by patient characteristics can better inform clinicians and patients when making treatment decisions.

Transportability methods have been developed and applied to extend the results of randomised trials to more representative patient populations. However, these methods typically require the use of individual participant data for either a single trial \citep{westreich_transportability_2017,dahabreh_generalizing_2019,dahabreh_learning_2024} or multiple trials in the meta-analysis setting \citep{dahabreh_toward_2020,dahabreh_efficient_2023}. Methods have also been developed to estimate the risk difference in a population of interest with differing baseline risk than was present in the trials \citep{murad_methods_2023}, but these methods require knowledge of risk in the general patient population eligible for treatment in the absence of treatment. 

Here, we consider the setting where the best available information to quantify the relative treatment effect is a summary measure from either a single trial or a meta-analysis of trials. Two sources of data are assumed to be available to us: 1) summary measures from a single trial or meta-analysis of trials quantifying the relative effect of treatment on the outcome of interest in the trial population(s), i.e., the risk ratio or hazard ratio, and 2) a representative sample from the general patient population who received treatment, with the outcome of interest and possibly other baseline covariates. Here it is assumed that uncontrolled confounding precludes estimation of treatment effects using general patient population data alone. 

Hereafter, the estimand is the average treatment effect on the treated (ATT) in the general patient population, defined as the difference in cumulative risk of the outcome for the treated patient population at a given time point under treatment versus no treatment. In our context, the ATT is an appropriate estimand for examining the absolute effect of endocrine therapy on breast cancer mortality because it quantifies the effect among patients who received treatment in routine practice, accounting for real-world eligibility criteria. In the described setting, neither data source can be used alone to identify this estimand. The trial is subject to selection bias and lacks external validity, while data from the general patient population are subject to uncontrolled confounding. 

This paper outlines identification assumptions sufficient to transport estimates of the risk ratio or hazard ratio from randomised trials to the treated patient population for estimation of the ATT. The implications of collapsibility of the relative effect measure for the validity of these identifiability conditions in practice are discussed. Estimators are proposed that are consistent for the ATT under the given assumptions, and their performance is evaluated through a series of simulation experiments. The approach is then applied to data from England's National Disease Registration Service to estimate the effect of endocrine therapy on risk of breast cancer death among women with oestrogen-receptor positive early breast cancer.

\section{Methods}\label{sec2:Methods}

\subsection{Preliminaries}
Let $A_i \in \{0,1\}$ represent a binary treatment assignment (e.g., endocrine therapy), and $T_i$ indicate time from treatment to the event of interest (e.g., breast cancer death) for participant $i$. The potential outcome under treatment assignment $a \in \{0,1\}$ is denoted by $T_i(a)$. The observed data may be right censored, such that the observed outcome is $T_i^*=\mathrm{min}(T_i,C_i)$, where $C_i$ indicates censoring time. Let $\delta_i=I(T_i^*=T_i)$ denote an observed event. Hereafter, subscripts $i$ will generally be omitted for notational ease.

Assume interest is in estimating the ATT for the general patient population $(R=0)$ at time $t$. This parameter can be expressed as $ATT_{R=0}(t)=\Pr(T(1)<t \mid A=1,R=0)-\Pr(T(0)<t \mid A=1,R=0)$. For example, we aim to estimate the effect of endocrine therapy on breast cancer mortality among oestrogen-receptor positive early breast cancer patients diagnosed between the ages of 20 and 50 in England between 2000-2009 who received endocrine therapy. 

To identify the ATT in the general patient population, consider two data sources. The first source of information is a summary measure of the relative treatment effect from one or more randomised trials. That is, assume one or more randomised controlled trials are conducted in their respective select patient populations, denoted by $R \in \{1, 2, ..., m\}$, where $m$ represents the number of randomised trials. In the single-trial setting ($m=1$), an estimate of the (intention to treat) relative treatment effect (i.e., risk ratio or hazard ratio) is available. When $m>1$, results from a meta-analysis based on these trials provide an estimate of the relative effect in a weighted average of their respective target populations. 

For the second information source, assume the treatment $A$ is subsequently applied in the general patient population $(R=0)$, where baseline covariates ($X$), and the outcome $T^*$ are also recorded. A representative sample from the treated population is available, with observed data $O_i=\{R_i=0, A_i=1,X_i,T_i^*,\delta_i\}$ for $i=1...n$, where $n$ represents the number of patients treated in the general patient population sample. Here, assume all patients enrolled in the trial(s) and general population patients are independent.

\subsection{Setting 1: Identifiability when the trial(s) estimate the risk ratio} \label{Sec.Methods.RR}
First consider the setting where the trial(s) provide estimates of the risk ratio. To transport the risk ratio from the trial population(s) to the treated patient population, additional assumptions are made about treatment effect heterogeneity. First, assume there is an absence of treatment effect heterogeneity for the risk ratio for all variables, both measured and unmeasured, that differ in distribution between the trial and treated patient populations. In the single-trial setting, the trial provides an estimate of the risk ratio in its trial population at a given time point $t$. When $m>1$, a meta-analysis provides an estimate of the common risk ratio across all trial populations at time point $t$. This trial population risk ratio is denoted by $RR_{R>0}(t)=\Pr(T(1)<t \mid  R>0) / \Pr(T(0)<t \mid R>0)$. Also consider the setting where there is treatment effect heterogeneity by discrete, measured baseline covariate(s) $X$, and the single trial or meta-analysis provides stratum-specific estimates of the risk ratio at a given time point $t$ for each level of $X$, i.e., $RR_{R>0}(t \mid X=x)=\Pr(T(1)<t \mid R>0,X=x) / \Pr(T(0)<t \mid R>0,X=x)$. In both settings, assume the trial(s) also provide estimates for corresponding measures of uncertainty, i.e.,  $SE(\widehat{RR}_{R>0}(t))$ in the absence of treatment effect heterogeneity or $SE(\widehat{RR}_{R>0}(t \mid X=x))$ for all $x$ when there is assumed to be treatment effect heterogeneity by $X$. Note these are typically provided on the log scale. Hereafter, assume that only this summary information is available from the trial(s).

Under the following assumptions $ATT_{R=0}(t)$ is identifiable from the observed data:

\begin{enumerate}
    \item No measurement error in the trial(s) and treated patient population: The treatment, outcome, and any covariates used in estimation are assumed to have been measured without error. \label{Assn:msmt}
    \item Causal consistency in the trial(s) and treated patient population: Assume $T= \sum_{a} I(A=a)T(a)$ and $C= \sum_{a} I(A=a)C(a)$ for $a \in \{0,1\}$,  where $C(a)$ is the potential censoring time under treatment $a$. \label{Assn:CC}
    \item Marginal exchangeability for treatment in the trial(s): $T(a) \perp A \mid R$ for $a \in \{0,1\}$ and $R>0$. 
    \item Positivity for treatment in the trial(s): $P(A=a \mid R)>0$ for $a \in \{0,1\}$ and $R>0$ 
	\item Independent censoring in the trial(s) and treated patient population: $T(a) \perp C(a) \mid A=a,R$ for $a \in \{0,1\}$ and $R \geq0$. \label{Assn:positivity}
    \item Transportability of the relative effect measure (see below for specific assumptions 6(a)-6(b) under marginal and conditional transportability of the risk ratio, respectively) \label{Assn:transp}
\end{enumerate}

Note that Assumptions 3 and 4 hold by design for well-conducted randomised trial(s). If the risk ratio is assumed to be marginally transportable from the trial(s) to the treated patient population, Assumption \ref{Assn:transp}(a) is:
\[
RR_{R>0}(t)
=
\frac{
  \Pr\{T(1)<t \mid A=1, R=0\}
}{
  \Pr\{T(0)<t \mid A=1, R=0\}
}\]

The identifiable form of $ATT_{R=0}(t)$ under assumption \ref{Assn:transp}(a), which includes data from the treated patient population and the transportable risk ratio from the trial population, is:
\begin{equation} \label{IDform.marg.RR}
ATT_{R=0}(t) = \Pr(T<t \mid A=1,R=0)
   \left[1-\{ RR_{R>0}(t) \}^{-1} \right]. 
\end{equation}

Alternatively, if the risk ratio is assumed to be transportable from the trial(s) to the treated patient population conditional on discrete baseline covariate(s) $X$, Assumption \ref{Assn:transp}(b) is: \[
\begin{aligned}
RR_{R>0}(t \mid X=x)
&=
\frac{
  \Pr\{T(1)<t \mid A=1, R=0, X=x\}
}{
  \Pr\{T(0)<t \mid A=1, R=0, X=x\}
} \\
&\quad \text{for all } x \text{ such that }
\Pr(X=x \mid A=1, R=0)>0.
\end{aligned}
\]                                                  

Under the stated assumptions, $ATT_{R=0}(t)$ is identifiable under Assumption \ref{Assn:transp}(b), with identifiable form:

\begin{equation}
\begin{split}
ATT_{R=0}(t)
&= \sum_x
\Pr(T<t \mid A=1, R=0, X=x) \\
&\quad {}\times
\left[1-\{RR_{R>0}(t \mid X=x)\}^{-1}\right]
\Pr(X=x \mid A=1, R=0).
\end{split}
\label{IDform.cond.RR}
\end{equation}     

The derivations of the identification results in Equations (\ref{IDform.marg.RR}) and (\ref{IDform.cond.RR}) are presented in Sections S.1 and S.2, respectively.

\subsection{Setting 2: Identifiability when the trial(s) estimate the causal hazard ratio}

To transport the hazard ratio from summary measures of a single trial or a meta-analysis of trials to the treated patient population, assume that proportional hazards holds up to time $t$  both in the trial population(s) and in the treated patient population. That is, if $S_a(t \mid R=r)=\Pr(T(a) \geq t \mid R=r)$ represents the survival function under treatment a in population $R=r$, then proportional hazards implies that $S_1(t \mid R=r)=\{S_0(t \mid R=r)\}^{HR_{R=r}}$, where $HR_{R=r}=h_1(t \mid R=r) / h_0(t \mid R=r)$ is the hazard ratio for population $R=r$ and $h_a(t \mid R=r)
=-\frac{\mathrm{d}}{\mathrm{d}t} \log\{S_a(t \mid R=r)\}$. Note for the general patient population, this assumption need only hold for the treated population such that $S_1(t \mid R=0,A=1)={S_0(t \mid R=0,A=1)}^{HR_{R=0,A=1}}$ where $HR_{R=0,A=1}=h_1(t \mid R=0,A=1)/h_0(t \mid R=0,A=1)$.

In the absence of treatment effect heterogeneity by measured or unmeasured covariates, a single trial provides an estimate of the hazard ratio for the trial population. When $m>1$, a meta-analysis provides an estimate of the hazard ratio across trial populations. For identification using a meta-analytic summary measure, we assume that the marginal hazard ratio is common across the trial populations, denoted $HR_{R>0}$. However, because of noncollapsibility, this assumption may not hold even in the absence of treatment effect heterogeneity if trial populations differ in their distributions of prognostic factors \citep{campbell2026hidden}. We return to this issue in Section \ref{sec2:Methods_collapsability}. When treatment effect heterogeneity by $X$ is assumed, summary measures from the trial(s) provide estimates of the hazard ratio for each level of $X$, i.e., $HR_{R>0,X=x}=h_1(t \mid  R>0,X=x)/h_0 (t \mid R>0,X=x)$. We similarly assume these stratum-specific hazard ratios are common across trial populations for identifiability. Assume that corresponding measures of uncertainty are provided such that estimates of $SE(\widehat{HR}_{R>0})$ in the absence of treatment effect heterogeneity or $SE(\widehat{HR}_{R>0,X=x})$ are available on the log scale for all $x$ when there is assumed to be treatment effect heterogeneity by $X$.

In these settings, $ATT_{R=0}(t)$ is identifiable from the observed data under the proportional hazards assumption described above, Assumptions \ref{Assn:msmt}– \ref{Assn:positivity} from Section \ref{Sec.Methods.RR}, and a revised version of Assumption \ref{Assn:transp} regarding transportability of the hazard ratio from the trial population(s) to the treated patient population. Specifically, if the hazard ratio is assumed to be marginally transportable from the trial(s) to the treated patient population, then Assumption \ref{Assn:transp}(c) is: \[
HR_{R>0}
=
HR_{R=0,A=1}
\]                                             
The identifiable form of $ATT_{R=0}(t)$ under assumption \ref{Assn:transp}(c) is:
\begin{equation} \label{IDform.marg.HR}
ATT_{R=0}(t) = \{S(t \mid A=1,R=0) \}^{HR_{R>0}^{-1}}
   - S(t \mid A=1,R=0). 
\end{equation} 
 
The marginal transportability assumption can be relaxed to assume the hazard ratio is transportable from the trial(s) to the treated patient population conditional on baseline covariate(s) $X$. Here, Assumption \ref{Assn:transp}(d) is: \[
\begin{aligned}
HR_{R>0,X=x}
&=
HR_{R=0,A=1,X=x}
=
\frac{
  h_1(t \mid A=1, R=0, X=x)
}{
  h_0(t \mid A=1, R=0, X=x)
} \\
&\quad \text{for all } x \text{ such that }
\Pr(X=x \mid A=1, R=0)>0.
\end{aligned}
\]   

Under the stated assumptions, $ATT_{R=0}(t)$ is identifiable under assumption \ref{Assn:transp}(d), with identifiable form:

\begin{equation}
\begin{split}
ATT_{R=0}(t)
&= \sum_x
\{S(t \mid A=1, R=0, X=x)^{HR_{R>0,X=x}^{-1}} - S(t \mid A=1, R=0, X=x) \} \\
&\quad {}\times
\Pr(X=x \mid A=1, R=0). \label{IDform.cond.HR}
\end{split}
\end{equation}     

The derivations of the identifiable forms (\ref{IDform.marg.HR}) and (\ref{IDform.cond.HR}) are presented in Sections S.3 and S.4, respectively. 

\subsection{Implications of collapsibility on Assumption \ref{Assn:transp}} \label{sec2:Methods_collapsability}

Whether or not the relative effect measure is collapsible has implications for the plausibility of assumptions \ref{Assn:transp}(a) - \ref{Assn:transp}(d) in real-world applications. The hazard ratio is a noncollapsible effect measure \citep{daniel2021making,didelez2022logic}.  
Consequently, even in the absence of treatment effect heterogeneity, marginal hazard ratios may differ between populations with different distributions of prognostic factors \citep{remiro2024transportability, phillippo2026multilevel,campbell2026hidden}. The magnitude of this difference depends on the distribution of prognostic factors and the strength of their associations with the outcome \citep{phillippo2026multilevel,campbell2026hidden}. This has implications both for estimating hazard ratios across multiple trial populations in the meta-analysis setting and for transporting hazard ratios from trials to the treated patient population. When making assumption \ref{Assn:transp}(c) or \ref{Assn:transp}(d), both treatment effect heterogeneity \textit{and} differences in distribution of prognostic factors between the trial population(s) and treated patient population along with the strength of these prognostic factors must be considered. In contrast, the risk ratio is a collapsible effect measure. When the risk ratio is homogeneous across populations either marginally or within covariate-defined strata, differences in distributions of prognostic factors do not themselves induce violations of assumptions \ref{Assn:transp}(a) or \ref{Assn:transp}(b).

\subsection{Estimation of the ATT}
\label{sec:Methods.est}

Plug-in estimators for $ATT_{R=0}(t)$ are proposed for the marginal and conditional transportability of the relative effect measures for both the settings where the risk ratio and the hazard ratio are assumed to be transportable from trial summary measures. These estimators are motivated by the identifiable forms of $ATT_{R=0}(t)$ in (\ref{IDform.marg.RR})-(\ref{IDform.cond.HR}).

For the marginal transportability setting, the estimators have the form:

\begin{equation}
    \widehat{ATT}_{R=0}^{(RR)}(t)=\widehat{\Pr}(T<t \mid A=1, R=0) \left [1-\{\widehat{RR}_{R>0}(t)\}^{-1} \right ]
    \label{EST.marg.RR}
\end{equation}

and

\begin{equation}
    \widehat{ATT}_{R=0}^{(HR)}(t)=\{\widehat{S}(t \mid A=1, R=0)\}^{\widehat{HR}_{R>0}^{-1}}-\widehat{S}(t \mid A=1, R=0)
    \label{EST.marg.HR}
\end{equation}

\noindent{when} the risk ratio and hazard ratio are transported, respectively. Here, $\widehat{\Pr}(T<t \mid A=1,R=0)$ or $\widehat{S}(t \mid A=1, R=0)$ are computed from the treated patient population using a consistent estimator of the cumulative risk of the outcome (or survival) at time $t$. In the absence of censoring and competing risks, these quantities can be the sample mean (or one minus the sample mean) at time $t$. In the presence of non-informative censoring, the Kaplan-Meier estimator can be used to compute these quantities. Here $\widehat{RR}_{R>0}(t)$ and $\widehat{HR}_{R>0}$ are consistent estimators of the common risk ratio or hazard ratio from the trial(s), respectively, obtained from summary measures from a single trial or meta-analysis. Under the previously defined assumptions and assuming $RR_{R>0}(t)$ and $HR_{R>0}$ are bounded away from zero, (\ref{EST.marg.RR}) and (\ref{EST.marg.HR}) are consistent estimators of $ATT_{R=0}(t)$ by the continuous mapping theorem.

For the conditional transportability setting, the plug-in estimators have the form:

\begin{equation}
\begin{split}
\widetilde{ATT}^{(RR)}_{R=0}(t)
&= \sum_x
\widehat{\Pr}(T<t \mid A=1, R=0, X=x) 
\left[1-\{\widehat{RR}_{R>0}(t \mid X=x)\}^{-1}\right]
\\
&\quad {}\times
\widehat{\Pr}(X=x \mid A=1, R=0)
\end{split}
\label{EST.cond.RR}
\end{equation}     and
\begin{equation}
\begin{split}
\widetilde{ATT}^{(HR)}_{R=0}(t)
&= \sum_x
\{\widehat{S}(t \mid A=1, R=0, X=x)^{\widehat{HR}_{R>0,X=x}^{-1}} - \widehat{S}(t \mid A=1, R=0, X=x) \} \\
&\quad {}\times
\widehat{\Pr}(X=x \mid A=1, R=0)
\end{split}
\label{EST.cond.HR}
\end{equation}     
when the covariate stratum-specific risk ratio and hazard ratio are transported, respectively. For these estimators, $\widehat{\Pr}(T<t \mid A=1, R=0, X=x)$, $\widehat{S}(t \mid A=1, R=0, X=x)$, and $\widehat{\Pr}(X=x \mid A=1, R=0)$ are computed from the treated patient population using consistent estimators of the 1) cumulative risk of the outcome at time $t$ for patients with $X=x$, 2) survival at time $t$ for patients with $X=x$, and 3) the proportion of patients in covariate stratum $X=x$, respectively. As in the marginal setting, quantity 1) can be the sample mean at time $t$ in the absence of censoring and competing events or the Kaplan-Meier estimator in the setting with non-informative censoring, and quantity 2) can be derived analogously. Quantity 3) can be computed as the proportion of patients in the sample in covariate stratum $X=x$. Here, $\widehat{RR}_{R>0}(t \mid X=x)$ and $\widehat{HR}_{R>0,X=x}$  are consistent estimators of the common risk ratios or hazard ratios from the trial(s) for patients with $X=x$, with estimates obtained from a single trial or meta-analysis. Similar to the marginal transportability setting, under suitable regularity conditions (\ref{EST.cond.RR}) and (\ref{EST.cond.HR}) are consistent estimators of $ATT_{R=0}(t)$.

When transporting hazard ratios using (\ref{EST.cond.HR}), this estimator relies on marginal estimates of the hazard ratio from the trial(s) for each covariate stratum. That is, it is assumed that marginal, stratum-specific hazard ratios were estimated for each level of $X$ rather than adjusted hazard ratios.

It is worth noting that analogous identification results can be derived along with corresponding estimators for the average treatment effect on the untreated (ATU) in the setting where relevant data sources are available and the identification assumptions are expected to hold. Such results would leverage the same trial summary measure(s) along with a representative sample from the general patient population who did not receive treatment to identify and estimate $ATU_{R=0}(t)=\Pr(T(1)<t \mid A=0,R=0)-\Pr(T(0)<t \mid A=0,R=0)$. 

\subsection{Variance estimation}
Variance of the proposed estimators (\ref{EST.marg.RR})-(\ref{EST.cond.HR}) must take into account not only variation in estimating quantities from the treated patient population but also variation in estimating the risk ratio or hazard ratio from the trial(s). Here, a Monte-Carlo procedure that accounts for both sources of variation is proposed. Specifically, $B$ independent, with replacement bootstrap resamples are selected from the treated patient population, where $B$ is a large number (e.g., 500 or more). 

In the marginal transportability setting, for each bootstrap resample $b$, a draw is taken from a normal distribution with mean $\log(\widehat{RR}_{R>0}(t))$ and standard deviation $\widehat{SE}(\log(\widehat{RR}_{R>0}(t)))$ for the risk ratio estimator or with mean $\log(\widehat{HR}_{R>0})$ and standard deviation $\widehat{SE}(\log(\widehat{HR}_{R>0}))$ for the hazard ratio estimator. The estimator (\ref{EST.marg.RR}) or (\ref{EST.marg.HR}) is applied to each bootstrap resample to obtain $\widehat{ATT}_{R=0,b}^{(RR)}$ or $\widehat{ATT}_{R=0,b}^{(HR)}$, and the standard error is computed as the standard deviation of the $B$ bootstrap estimates. Wald-type confidence intervals can be constructed from point and variance estimates. Alternatively, percentile-based confidence intervals can be derived from the bootstrap estimates. Note that Monte-Carlo draws from the risk and hazard ratio are taken on the log scale as the log transformed values are expected to be normally distributed, but not necessarily the untransformed risk or hazard ratios themselves. 

Variance is estimated analogously in the conditional transportability setting, but with covariate-stratum specific draws from the relative effect measure distributions and application of estimators (\ref{EST.cond.RR}) or (\ref{EST.cond.HR}) to each bootstrap resample.

\section{Simulations}\label{sec3:Sims}

\subsection{Simulation setup}
A simulation study was conducted to examine the finite sample performance of the proposed estimators under a variety of data-generating mechanisms. The estimators were applied in scenarios with no treatment effect heterogeneity and in scenarios with treatment effect heterogeneity by observed or unobserved covariates. To better reflect real-world settings, the distributions of prognostic factors differed across trials (in the meta-analysis setting) and between the trial population(s) and treated patient population. Baseline risk or hazard also varied between the trial population(s) and the patient population. As discussed in Section \ref{sec2:Methods_collapsability}, such differences can result in violations of assumptions \ref{Assn:transp}(c) or \ref{Assn:transp}(d) due to noncollapsibility of the hazard ratio, even in the absence of treatment effect heterogeneity. Simulations considered both moderate and strong prognostic effects, as larger departures from hazard ratio transportability were anticipated when prognostic effects were stronger. Differences in the strength of prognostic factors alone were not expected to affect the performance of the risk ratio estimators. Simulations and data analysis were conducted using R version 4.5.1 (Vienna, Austria).

In the primary simulations, data were simulated for $15$ trials, each of size $650$. In each trial, a covariate $X$ (measured) was simulated from a Bernoulli distribution with probability $\theta$ where $\theta$ was drawn independently for each of the $15$ trials from a uniform distribution ranging between $0.08$ and $0.20$. This created differences in the distribution of the prognostic factor $X$ across trial populations. An additional covariate $U$ (unmeasured) was simulated from a Bernoulli distribution with probability $0.08$ in all trial populations. Treatment was randomly assigned 1:1. 

To examine performance of the estimators when either the risk ratio or hazard ratio was assumed to be transportable, outcomes were generated two ways. First, for the setting where the risk ratio was assumed to be transportable, the probability of the outcome (e.g., death) by 15 years under each treatment $a \in \{0,1\}$ was simulated from a Bernoulli distribution with probability $\mu_a= \exp\{\beta_0+\beta_XX+\beta_UU+z\}$ , where $\beta_0=\log(0.15)$ and $z$ is specified in Table \ref{tab1} for the five simulation scenarios considered. For the setting with moderate prognostic factors, $\beta_X=\log(1.8)$ and $\beta_U=\log(1.6)$. For the setting with strong prognostic factors,  $\beta_X=\log(2.5)$ and $\beta_U=\log(2.2)$. Then, causal consistency was applied to obtain the observed outcome for each observation, such that the untreated and treated were assigned binary outcomes based on random draws from $\mu_0$ and $\mu_1$, respectively.  

\begin{table}
\begin{threeparttable}
\caption{Specifications of simulation parameters by scenario, transporting the risk ratio}
\label{tab1}
\centering
\small
\setlength{\tabcolsep}{3.5pt} 

\begin{tabular}{c l c c c} 
\hline
\multirow{3}{*}{Scenario} &
\multirow{3}{*}{Description} &
\multirow{3}{*}{$z$} &
\multicolumn{2}{c}{$ATT_{R=0}(t=15)$} \\
\cline{4-5}
& & & \makecell{Moderate \\ prognostic} & \makecell{Strong \\ prognostic} \\
\hline
1A & MT & $a \log(0.9)$                                  & -0.025 & -0.035 \\
1B & CT by $X$, minor & $a \log(0.85)X+a \log(0.9)(1-X)$ & -0.033 & -0.047 \\
1C & CT by $X$, major & $a \log(0.75)X+a \log(0.9)(1-X)$ & -0.047 & -0.070 \\
1D & CT by $U$, minor & $a \log(0.85)U+a \log(0.9)(1-U)$ & -0.030 & -0.044 \\
1E & CT by $U$, major & $a \log(0.75)U+a \log(0.9)(1-U)$ & -0.040 & -0.061 \\
\hline
\end{tabular}
\begin{tablenotes}
      \small
    \item Note: Minor and major refer to the degree of heterogeneity in the risk ratio by $X$. Moderate and strong prognostic refer to the strength of prognostic factors $X$ and $U$, i.e., $\beta_X=\log(1.8)$ and $\beta_U=\log(1.6)$ for the moderate prognostic setting and $\beta_X=\log(2.5)$ and $\beta_U=\log(2.2)$ for the strong prognostic setting. MT = marginal transportability; CT = conditional transportability.
    \end{tablenotes}
      \end{threeparttable}
\end{table}
 
For the setting where the hazard ratio is assumed to be transportable, event times were generated under a proportional hazards model. For each treatment $a$, the hazard function was assumed to be $h_a(t \mid R>0,X,U)=\lambda \exp(\gamma X + \nu U + \eta a + \phi aX + \xi aU)$, where $\lambda=0.0267$ (the baseline hazard in the trial populations). The parameters $\gamma$ and $\nu$ represent prognostic effects for $X$ and $U$, respectively. For the setting with moderate prognostic factors, $\gamma=\nu=0.4$. For the setting with strong prognostic factors, $\gamma=\nu=1.0$. Thus, the corresponding hazard ratios for the prognostic variables were approximately 1.5 in the moderate setting and 2.7 in the strong setting. The remaining parameter values control the treatment effect and are specified in Table \ref{tab2} for each scenario. Event times and censoring times were generated using inverse-transform sampling from exponential distributions under each treatment for each scenario, with follow-up administratively censored at 15 years, as follows:

\begin{equation}
   T(a)= - \frac{\log(N)}{\lambda \exp(\gamma X+\nu U+\eta a+ \phi aX+ \xi aU)} 
   \label{sim.surv.times}
\end{equation}

\begin{equation}
   C(a)=\min\{15,(-\log(M))/({\log(0.92)/15})\}  
   \label{sim.cens.times}
\end{equation}

\noindent{where} $N$ and $M$ are independent uniform numbers between 0 and 1. This data generating process resulted in approximately $8\%$ random dropout by year 15 with the remaining units that were event free at 15 years administratively censored. The final observed event time for each unit was  $T_i^*=\min(T_i, C_i)$, where $T_i=I(A_i=1)T(1)+I(A_i=0)T(0)$ and $C_i=I(A_i=1)C(1)+I(A_i=0)C(0)$. 

\begin{table}
  \begin{threeparttable}  
\caption{Specifications of simulation parameters by scenario, transporting the hazard ratio}
\label{tab2}
\centering
\small
\setlength{\tabcolsep}{3.5pt} 

\begin{tabular}{c l c c c c c} 
\hline
\multirow{3}{*}{Scenario} &
\multirow{3}{*}{Description} &
\multirow{3}{*}{$\eta$} &
\multirow{3}{*}{$\phi$} &
\multirow{3}{*}{$\xi$} &
\multicolumn{2}{c}{$ATT_{R=0}(t=15)$} \\
\cline{6-7}
& & & & & \makecell{Moderate \\ prognostic} & \makecell{Strong \\ prognostic} \\
  \hline 
2A	&	MT	&	$\log(0.70)$	& 0 & 0 & 	-0.107 & -0.106 \\
2B	&	CT by $X$, minor 	&	$\log(0.71)$	& $\log(0.63)-\log(0.71)$ & 0 & -0.119	& -0.119	 \\
2C	&	CT by $X$, major 	& $\log(0.80)$ & $\log(0.55)-\log(0.80)$ & 0 & -0.117	& -0.119 \\
2D	&	CT by $U$, minor 	&	$\log(0.71)$	& 0 & $\log(0.63)-\log(0.71)$	& -0.114 & -0.113 \\
2E	&	CT by $U$, major 	&	$\log(0.80)$	& 0 & $\log(0.55)-\log(0.80)$ & -0.103	& -0.100 \\
\hline
\end{tabular}
\begin{tablenotes}
      \small
    \item Note: Minor and major refer to the degree of heterogeneity in the hazard ratio by $X$. Moderate and strong prognostic refer to the strength of prognostic factors $X$ and $U$, i.e., $\gamma=\nu=0.4$ for the moderate prognostic setting and $\gamma=\nu=1.0$ for the strong prognostic setting.  MT = marginal transportability; CT = conditional transportability.
    \end{tablenotes}
    \end{threeparttable}
\end{table}

A patient population of size $15,000$ was also simulated, where $X$ and $U$ were Bernoulli random variables with probabilities $0.4$ and $0.3$, respectively. Treatment $A$ was Bernoulli with probability $\operatorname{expit}(1.2+0.6X-0.24U)$. For Scenarios 1A-1E, potential outcomes were simulated from a Bernoulli for $a \in \{0,1\}$ with probability equal to $\mu_a= \exp(\beta_0+\beta_XX+\beta_UU+z)$ with $\beta_0=\log(0.16)$, $z$ as specified in Table \ref{tab1}, and $\beta_X$ and $\beta_U$ specified as in the trial populations. In Scenarios 2A-2E, potential outcomes were simulated from (\ref{sim.surv.times}) and (\ref{sim.cens.times}), with $\lambda=0.03$, $\gamma$ and $\nu$ specified as in the trial populations, and the remaining parameter values specified in Table \ref{tab2} for each scenario. Causal consistency was applied to assign observed outcomes based on treatment.

In Scenario 1A, the risk ratio is marginally transportable from the trial population to the treated patient population, so both assumptions \ref{Assn:transp}(a) and \ref{Assn:transp}(b) hold. In Scenarios 1B and 1C, the risk ratio is heterogeneous by $X$, so the marginal transportability assumption \ref{Assn:transp}(a) is violated but the conditional transportability assumption \ref{Assn:transp}(b) holds. In Scenarios 1D and 1E, the risk ratio is heterogeneous by $U$, an unmeasured covariate which differs in distribution between the trial and patient populations, so both the marginal and conditional transportability assumptions are violated. Effect heterogeneity is stronger in scenarios 1C and 1E than in scenarios 1B and 1D. Scenarios are defined analogously for the hazard ratio setting (Scenarios 2A-2E). However, due to noncollapsibility of the hazard ratio, differing distributions of prognostic factors between trial population(s) and the treated patient population are expected to lead to violations of assumptions \ref{Assn:transp}(c) and \ref{Assn:transp}(d) in all scenarios. Performance of hazard ratio estimators is expected to be better in the setting where prognostic effects are moderate compared to the setting where prognostic effects are strong.

For Scenarios 1A-1E, risk ratios were estimated from the trials using the Mantel-Haenszel fixed-effects meta-analysis approach in the “meta” package in R \citep{balduzzi_how_2019}, both overall (for the marginal estimator) and stratified by $X$ (for the conditional estimator). The estimators (\ref{EST.marg.RR}) and (\ref{EST.cond.RR}) were applied to data from each scenario for estimation of the ATT at 15 years. Cumulative risk of the outcome was estimated based on a sample mean of the observed outcomes. 

For Scenarios 2A-2E, the hazard ratio was estimated for each trial using a Cox proportional hazards regression model. Then, Cox model results were pooled across trials using the metagen() function in the “meta” package in R, which pools estimates using an inverse variance weighting approach. The marginal estimator (\ref{EST.marg.HR}) was applied for estimation of the ATT. For the conditional estimator, the hazard ratio was estimated separately in each trial for units with $X=0$ and $X=1$, and the meta-analysis separately pooled hazard ratios for each level of $X$. The conditional  estimator (\ref{EST.cond.HR}) was then applied to data from each scenario. For both estimators, survival in the treated patient population was estimated based on the Kaplan-Meier estimator.

The variance in each scenario was estimated using the proposed Monte-Carlo procedures with 500 replications. To demonstrate the consequences of ignoring variation in the meta-analysis estimates, variances were also estimated for each scenario assuming that the trial risk ratios or hazard ratios were measured without error, rather than selecting draws from Normal distributions within each bootstrap sample.

To examine performance of the methods in the single-trial setting, additional simulations were conducted using the same data generating mechanisms described above, but with $\theta=0.20$ for the single trial and with trial sample sizes of $650$, $1500$, and $5000$ (to examine finite sample bias). Marginal risk ratios and hazard ratios were estimated in each trial overall and for each level of $X$. 

For each scenario, 2000 simulations were conducted. Bias, relative bias (bias divided by the true ATT), empirical standard error, average standard error, standard error ratios, and empirical 95\% confidence interval coverage were computed for each estimator. The true values of the ATT for each scenario were determined empirically by averaging $\Pr(T(1)<t \mid A=1,R=0)-\Pr(T(0) <t \mid A=1,R=0)$ from these data generating mechanisms across 50 million realisations from the patient population (Tables \ref{tab1} and \ref{tab2}). 

\subsection{Simulation results}

\subsubsection{Transporting summary measures from a meta-analysis}
\begin{figure} [htbp]
\begin{center}
  \includegraphics[
  width=1.00\columnwidth,
  keepaspectratio,
  trim=0cm 2cm 0cm 0cm,
  clip
]{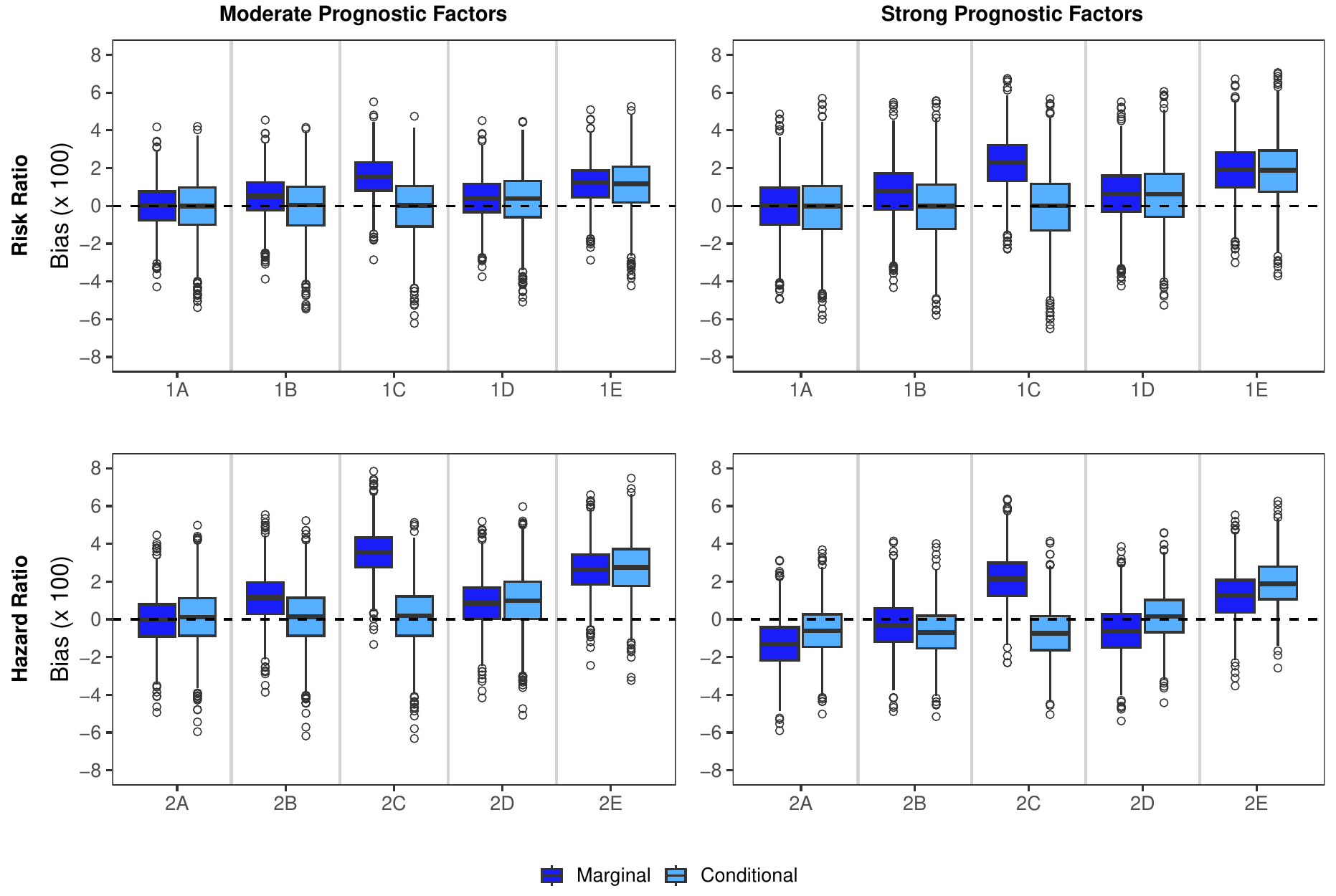}
  \caption{Empirical bias when transporting summary measures from a meta-analysis of 15 trials by simulation scenario for marginal and conditional estimators of the ATT at $t=15$, 2000 simulations. Boxplots show the median (horizontal line), interquartile range (box), whiskers extending to 1.5 times the interquartile range beyond the quartiles, and outliers (open circles). Moderate and strong prognostic factors refer to the strength of the prognostic variables $X$ and $U$. For the risk ratio scenarios 1A-1E, $\beta_X=\log(1.8)$ and $\beta_U=\log(1.6)$ for the moderate prognostic setting and $\beta_X=\log(2.5)$ and $\beta_U=\log(2.2)$ for the strong prognostic setting. For the hazard ratio scenarios 2A-2E, $\gamma=\nu=0.4$ for the moderate prognostic setting and $\gamma=\nu=1.0$ for the strong prognostic setting.}
  \label{fig1}
\end{center}
\end{figure}

The estimators performed as expected in simulations, with results presented in Table \ref{tab3}, Figure \ref{fig1}, and Table S1. In Scenario 1A, when marginal transportability of the risk ratio to the treated patient population held, both marginal and conditional estimators were empirically unbiased (Table \ref{tab3}, Figure \ref{fig1}). In Scenarios 1B and 1C, when marginal transportability was violated but conditional transportability held, the conditional estimator (\ref{EST.cond.RR}) was empirically unbiased but the marginal estimator (\ref{EST.marg.RR}) was biased. Bias was larger in Scenarios 1C compared to 1B, as effect heterogeneity by $X$ was stronger. Both marginal and conditional estimators were biased in Scenarios 1D and 1E, when there was effect heterogeneity by an unmeasured variable $U$, with larger bias when there was stronger effect heterogeneity (Scenario 1E). Due to collapsibility of the risk ratio, performance of the estimators (\ref{EST.marg.RR}) and (\ref{EST.cond.RR}) was similar when prognostic factors were moderate and strong. 

When prognostic effects were moderate, departures from the transportability of the hazard ratio assumption were relatively small, and empirical bias for the estimators (\ref{EST.marg.HR}) and (\ref{EST.cond.HR}) was low when treatment effect heterogeneity was appropriately taken into account. That is, empirical bias was low for the marginal estimator (\ref{EST.marg.HR}) in Scenario 2A and for the conditional estimator (\ref{EST.cond.HR}) in Scenarios 2B and 2C. Both estimators were biased in Scenarios 2D and 2E when there was treatment effect heterogeneity that was not accounted for by the estimator. However, when prognostic effects were strong, both estimators exhibited bias in all scenarios. That is, even when there was no treatment effect heterogeneity, differences in the distributions of prognostic factors between populations implied that $HR_{R>0} \neq HR_{R=0,A=1}$. This was most notable in Scenario 2A, where there was no treatment effect heterogeneity but relative bias increased from 0.5\% in the moderate setting to 12.2\% in the setting where prognostic factors were strong. This demonstrates how the validity of Assumptions \ref{Assn:transp}(c) and \ref{Assn:transp}(d) is dependent on the presence of prognostic factors that differ in distribution between the trial population(s) and the treated patient population. The magnitude of bias under violation of these assumptions depends in part on the strength of the effect of the prognostic factors on the outcome and the magnitude of differences in the distribution of the prognostic factors between populations.

\begin{table} 
\begin{threeparttable}
\caption{Simulation summary results, transporting summary measures from a meta-analysis of 15 trials, 2000 simulations}
\label{tab3}
\centering
\small
\setlength{\tabcolsep}{3.5pt} 

\begin{tabular}{l c l c c c c c c} 
\hline
  \makecell{Prognostic \\ Factors}  & Scenario & Estimator & \makecell{Bias \\ (x100)} & \makecell{Relative \\ Bias(\%)} & ASE & ESE & SER & \makecell{95\% CI \\ Coverage} \\ 
  \hline 
Moderate & 1A & Marginal & 0.0 & 0.5 & 1.15 & 1.13 & 1.02 & 96 \\
 & 1A & Conditional & 0.0 & 1.8 & 1.44 & 1.47 & 0.98 & 94 \\
 & 1B & Marginal & 0.5 & -15.2 & 1.13 & 1.11 & 1.02 & 92 \\
 & 1B & Conditional & 0.0 & 1.0 & 1.46 & 1.49 & 0.98 & 95 \\
 & 1C & Marginal & 1.5 & -32.3 & 1.10 & 1.07 & 1.02 & 69 \\
 & 1C & Conditional & 0.0 & 1.0 & 1.53 & 1.55 & 0.99 & 95 \\
 & 1D & Marginal & 0.4 & -12.7 & 1.14 & 1.11 & 1.02 & 94 \\
 & 1D & Conditional & 0.4 & -11.6 & 1.42 & 1.45 & 0.98 & 92 \\
 & 1E & Marginal & 1.2 & -28.9 & 1.10 & 1.08 & 1.02 & 80 \\
 & 1E & Conditional & 1.1 & -28 & 1.38 & 1.42 & 0.98 & 83 \\
 \hline
Strong & 1A & Marginal & 0.0 & 0.8 & 1.49 & 1.47 & 1.02 & 95 \\
 & 1A & Conditional & -0.1 & 2.2 & 1.71 & 1.71 & 1.00 & 95 \\
 & 1B & Marginal & 0.7 & -15.6 & 1.47 & 1.45 & 1.02 & 91 \\
 & 1B & Conditional & 0.0 & 1.0 & 1.75 & 1.75 & 1.00 & 95 \\
 & 1C & Marginal & 2.2 & -32.0 & 1.42 & 1.40 & 1.01 & 63 \\
 & 1C & Conditional & -0.1 & 1.0 & 1.84 & 1.86 & 0.99 & 95 \\
 & 1D & Marginal & 0.6 & -14.1 & 1.47 & 1.44 & 1.02 & 92 \\
 & 1D & Conditional & 0.6 & -13.0 & 1.69 & 1.68 & 1.01 & 92 \\
 & 1E & Marginal & 1.9 & -31.2 & 1.43 & 1.39 & 1.02 & 73 \\
 & 1E & Conditional & 1.8 & -30.3 & 1.64 & 1.63 & 1.01 & 77 \\
\hline
 Moderate & 2A & Marginal & 0.0 & 0.5 & 1.24 & 1.26 & 0.99 & 95 \\
 & 2A & Conditional & 0.1 & -0.8 & 1.53 & 1.54 & 0.99 & 94 \\
 & 2B & Marginal & 1.1 & -9.5 & 1.23 & 1.25 & 0.99 & 84 \\
 & 2B & Conditional & 0.1 & -0.9 & 1.56 & 1.57 & 0.99 & 94 \\
 & 2C & Marginal & 3.6 & -30.3 & 1.17 & 1.18 & 0.98 & 15 \\
 & 2C & Conditional & 0.1 & -1.2 & 1.59 & 1.60 & 1.00 & 95 \\
 & 2D & Marginal & 0.8 & -7.4 & 1.23 & 1.25 & 0.99 & 89 \\
 & 2D & Conditional & 1.0 & -8.5 & 1.52 & 1.53 & 0.99 & 89 \\
 & 2E & Marginal & 2.6 & -25.7 & 1.17 & 1.19 & 0.98 & 38 \\
 & 2E & Conditional & 2.7 & -26.4 & 1.46 & 1.49 & 0.98 & 53 \\
 \hline
 Strong & 2A & Marginal & -1.3 & 12.2 & 1.28 & 1.30 & 0.99 & 83 \\
 & 2A & Conditional & -0.6 & 5.6 & 1.26 & 1.28 & 0.98 & 91 \\
 & 2B & Marginal & -0.3 & 2.7 & 1.28 & 1.30 & 0.99 & 94 \\
 & 2B & Conditional & -0.7 & 5.7 & 1.27 & 1.29 & 0.98 & 91 \\
 & 2C & Marginal & 2.1 & -17.9 & 1.26 & 1.28 & 0.98 & 59 \\
 & 2C & Conditional & -0.7 & 6.1 & 1.27 & 1.30 & 0.98 & 90 \\
 & 2D & Marginal & -0.6 & 5.4 & 1.28 & 1.29 & 0.99 & 92 \\
 & 2D & Conditional & 0.2 & -1.4 & 1.26 & 1.29 & 0.98 & 94 \\
 & 2E & Marginal & 1.2 & -12.4 & 1.25 & 1.26 & 0.99 & 83 \\
 & 2E & Conditional & 1.9 & -19.1 & 1.24 & 1.27 & 0.97 & 66 \\
\hline
\end{tabular}
\begin{tablenotes}
      \small
      \item Note: Bias, relative bias, ASE, ESE, SER, and 95\% CI coverage are calculated for the ATT at $t=15$. ASE, SER, and 95\% CI coverage are based on the Monte-Carlo procedure with 500 replications. Moderate and strong prognostic factors refer to the strength of the prognostic variables $X$ and $U$. For the risk ratio scenarios 1A-1E, $\beta_X=\log(1.8)$ and $\beta_U=\log(1.6)$ for the moderate prognostic setting and $\beta_X=\log(2.5)$ and $\beta_U=\log(2.2)$ for the strong prognostic setting. For the hazard ratio scenarios 2A-2E, $\gamma=\nu=0.4$ for the moderate prognostic setting and $\gamma=\nu=1.0$ for the strong prognostic setting.
Abbreviations: ASE = Average standard error; ESE = empirical standard error; SER = standard error ratio; CI = confidence interval.

    \end{tablenotes}
    \end{threeparttable}
\end{table}

The proposed Monte-Carlo procedure for estimating the variance performed well, with the empirical standard error tracking with the average standard error, leading to standard error ratios close to one for all scenarios, marginal and conditional estimators, and for both effect measures (Table \ref{tab3}). For settings where the estimators were empirically unbiased, this led to approximately nominal 95\% confidence interval coverage. In Scenario 1A, when marginal transportability of the risk ratio held, the marginal estimator (\ref{EST.marg.RR}) was more efficient than the conditional estimator (\ref{EST.cond.RR}). In scenarios where point estimates were biased, confidence interval coverage was generally below the nominal level. 

When variation in estimation of the risk ratio or hazard ratio from the meta-analysis was ignored, the resulting variance estimators were substantially biased, as expected. This led to confidence interval coverage well below the nominal level, i.e., $\leq 12\%$ in all settings and standard error ratios $<0.08$ for all scenarios (Table S1). 

\subsubsection{Transporting summary measures from a single trial}
When the single trial sample size was large enough to yield precise summary measures (e.g., 5000), the estimators performed similarly in the single-trial setting as in the meta-analysis setting (Table S4). However, when the trial sample size was small (e.g., 650), the single trial summary measures were much less precise than in the meta-analysis setting, resulting in substantial finite sample bias (Table S2). This bias diminished as the trial sample size increased to 1500 (Table S3) and was negligible at a sample size of 5000 (Table S4). The variance estimator performed reasonably well even when single-trial estimates were imprecise. Even for smaller sample sizes, confidence interval coverage remained close to the nominal level in scenarios where the estimators were expected to be consistent.

\section{Application}\label{sec4:App}
The Early Breast Cancer Trialists’ Collaborative Group (EBCTCG) conducted a meta-analysis to estimate the effect of five years of adjuvant tamoxifen on breast cancer outcomes, including breast cancer mortality \citep{early_breast_cancer_trialists_collaborative_group_ebctcg_relevance_2011}. Tamoxifen is a common endocrine therapy treatment for patients with oestrogen-receptor positive breast cancer. The meta-analysis was based on individual patient data from 21,457 patients in 20 randomised trials. For patients with oestrogen-receptor positive breast cancer, the estimated hazard ratio for five-year tamoxifen versus no tamoxifen on breast cancer mortality over the first 15 years of follow-up was 0.68 (95\% CI: 0.63-0.74). This hazard ratio was estimated by combining log-rank statistics calculated within strata defined by trial, baseline prognostic factors, and years since randomisation.  The estimated hazard ratio was similar across patient age, nodal status, tumour grade and size, and chemotherapy use.

 The estimator (\ref{EST.marg.HR}) from Section \ref{sec:Methods.est} was applied to the hazard ratio reported by the EBCTCG meta-analysis and data from England's National Disease Registration Service to estimate the effect of adjuvant endocrine therapy on cumulative risk of breast cancer death at 15 years. That is, for the purposes of this application, the EBCTCG estimate was assumed to equal the marginal hazard ratio in the treated patient population as under Assumption 6(c). The construction of the cohort based on National Disease Registration Service data has been previously described \citep{taylor_breast_2023}. The treated patient cohort included patients in England with an initial diagnosis of early breast cancer between 2000-2009 aged 20-50 at the time of diagnosis who likely received endocrine therapy. These restrictions allowed for 15 years of follow-up and limited to primarily premenopausal women to better align with the population of women recommended for tamoxifen. Exclusions to the cohort are indicated in Figure S1. Briefly, those with characteristics that would exclude them from trials, including additional cancer diagnoses, those not receiving surgery, and those with neoadjuvant treatments, were excluded. In addition, those with less than three months of follow-up (n=510) were also excluded. To coincide approximately with the timing of treatment initiation, patients were followed from three months after initial breast cancer diagnosis until death, with administrative censoring at 15 years.

Table \ref{tab2.Supp} presents patient and tumour-level characteristics of the treated patients in the cohort ($n=22,726$). These characteristics were measured prior to the receipt of endocrine therapy. The majority of patients were over age 45 (53.4\%), white (89.3\%), and had no known comorbidities (97.6\%). Approximately half of patients (48.4\%) had small tumours (1-20 mm). A substantial proportion of patients (45.8\%) had unknown nodal involvement. Among those for which the number of positive nodes was recorded, $55\%$ had one or more positive nodes. Approximately 40\% of patients had recorded chemotherapy prior to endocrine therapy.

\begin{table}
\begin{threeparttable}
\caption{Characteristics of the cohort}
\label{tab2.Supp}
\centering
\small
\setlength{\tabcolsep}{3.5pt} 

\begin{tabular}{l l c} 
\hline
  Characteristic &   & $N(\%)$ \\ 
  \hline 
Age at diagnosis (years)	&	20-44	&	10,601 (46.6)	\\
	&	45-50	&	12,125 (53.4)	\\
Index of Multiple Deprivation	&	$<$20\% - least deprived	&	5,201 (22.9)	\\
	&	20-39\%	&	5,084 (22.4)	\\
	&	40-59\%	&	4,732 (20.8)	\\
	&	60-79\%	&	4,257 (18.7)	\\
	&	80\%+ - most deprived	&	3,452 (15.2)	\\
Geographical region	&	Eastern	&	3,959 (17.4)	\\
	&	North West	&	1,048 (4.6)	\\
	&	Northern and Yorkshire	&	5,351 (23.5)	\\
	&	Oxford	&	  442 (1.9)	\\
	&	South West	&	1,074 (4.7)	\\
	&	Thames	&	5,285 (23.3)	\\
	&	Trent	&	1,130 (5.0)	\\
	&	West Midlands	&	4,437 (19.5)	\\
Ethnicity	&	White	&	20,295 (89.3)	\\
	&	Asian or Asian British	&	   692 (3.0)	\\
	&	Black or Black British	&	   516 (2.3)	\\
	&	Other Ethnic Groups	&	   330 (1.5)	\\
	&	Mixed	&	   110 (0.5)	\\
	&	Unknown	&	   783 (3.4)	\\
Charlson comorbidity index	&	0	&	22,182 (97.6)	\\
	&	1	&	   448 (2.0)	\\
	&	2+	&	    96 (0.4)	\\
Tumour size	&	1-20 mm	&	10,989 (48.4)	\\
	&	21-50 mm	&	 7,499 (33.0)	\\
	&	$>$50 mm	&	   935 (4.1)	\\
	&	Unknown	&	 3,303 (14.5)	\\
Number of positive nodes	&	0	&	 5,599 (24.6)	\\
	&	1 to 3	&	 4,546 (20.0)	\\
	&	4 to 9	&	 1,562 (6.9)	\\
	&	10 or more	&	   608 (2.7)	\\
	&	Unknown	&	10,411 (45.8)	\\
Chemotherapy $\dagger$	&	Recorded	&	 9,198 (40.5)	\\
	&	Not recorded	&	13,528 (59.5)	\\
\hline
Total	&		&	22,726 (100)	\\
\hline
\end{tabular}
\begin{tablenotes}
      \small
      \item $\dagger$ Based on treatment guidelines in England, chemotherapy is given prior to endocrine therapy. 
    \end{tablenotes}
    \end{threeparttable}
\end{table}

\begin{figure}[htbp]
\centering
\includegraphics[
  width=0.98\columnwidth,
  height=0.85\textheight,
  keepaspectratio,
  trim=4cm 18cm 4cm 18cm,
  clip
]{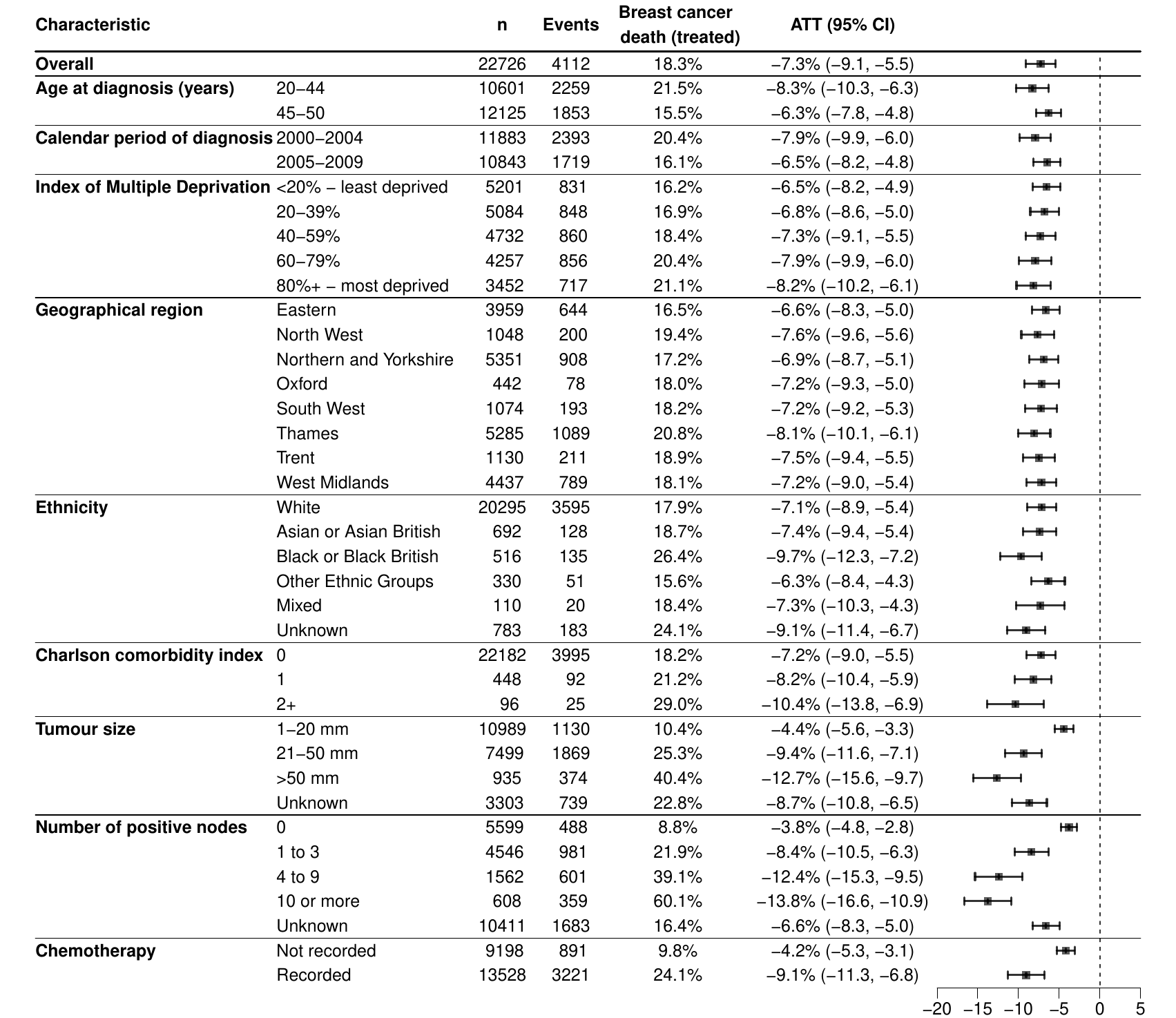}
\caption{Estimated average treatment effect on the treated (ATT) of endocrine therapy on cumulative risk of breast cancer death at 15 years, overall and across patient subgroups, among women diagnosed with early breast cancer in England during 2000–2009 at ages 20–50 years. Note: Breast cancer death (treated) is the estimated 15-year cumulative risk of breast cancer death among patients receiving endocrine therapy. Negative ATT values indicate a lower 15-year cumulative risk of breast cancer death with endocrine therapy compared with no endocrine therapy.}
\label{fig2}
\end{figure}

In the treated patient cohort, breast cancer survival at 15 years was estimated based on the Kaplan-Meier estimator. That is, for the purposes of this analysis, non-breast cancer deaths were censored. The hazard ratio from the EBCTCG meta-analysis was marginally transported to the treated patient population to estimate the ATT of endocrine therapy versus no endocrine therapy on the cumulative risk of breast cancer death at 15 years by patient and tumour characteristics. Variances were estimated using the proposed Monte-Carlo procedure with $2500$ replications. The results are presented in Figure \ref{fig2}. 

In the treated patient cohort, the estimated 15-year cumulative risk of breast cancer death was 18.3\%. The overall estimated ATT in the treated patient population was -7.3\% (95\% CI: -9.1\%, -5.5\%). That is, based on the marginal transportability of the hazard ratio assumption, women treated with endocrine therapy are estimated to have experienced an absolute 7.3\% lower risk of breast cancer death at 15 years compared to  non-receipt of endocrine therapy. The estimated benefit of endocrine therapy on breast cancer survival was larger for patients with higher risk tumours (e.g., larger tumours or more positive nodes) and smaller for women whose breast cancer was detected by screening. For example, among women with small tumours (1-20mm), endocrine therapy was estimated to reduce the risk of death by 4.4\% (95\% CI: 3.3\%, 5.6\%). Those with large tumours ($>$50mm) were estimated to have a much greater benefit from endocrine therapy, with an ATT of -12.7\% (95\% CI: -15.6\%, -9.7\%).  These results demonstrate that while the relative effect of endocrine therapy on cumulative risk of breast cancer death might be assumed constant across patient and tumour characteristics, absolute effects vary based on differences in baseline risk for different patient populations. 

While this example provides a demonstration of the methods described in Section \ref{sec2:Methods}, these results must be interpreted with caution because application of estimator (\ref{EST.marg.HR}) requires the EBCTCG reported hazard ratio to represent the marginal hazard ratio as specified in Assumption 6(c). Even if the EBCTCG estimate is taken to represent the marginal hazard ratio, as previously described and demonstrated in the simulation study, the marginal hazard ratio estimator (\ref{EST.marg.HR}) from Section \ref{sec:Methods.est} can be biased even in the absence of treatment effect heterogeneity if there are strong prognostic factors that differ in distribution between the trial population(s) and the treated patient population. To examine the potential for such bias, unadjusted associational hazard ratios for breast cancer mortality were estimated for each available prognostic factor with data from the treated patient cohort using univariable Cox proportional hazards models (Table S5). Three prognostic factors had large associational hazard ratios ($>2$): tumour size, the number of positive nodes, and recorded chemotherapy. While EBCTCG did not provide the distribution of participant characteristics in the meta-analysis, the distribution of oestrogen-receptor positive trial participants can be estimated from Figure 3 in \citep{early_breast_cancer_trialists_collaborative_group_ebctcg_relevance_2011} for two of the strong prognostic factors. In EBCTCG trials, $44\%$ of oestrogen-receptor positive participants had one or more positive nodes and $51\%$ received chemotherapy. These distributions can be compared with the treated patient population characteristics in Table \ref{tab2.Supp}. While incomplete recording of nodal status for the treated patient population limits direct comparisons, there is evidence that the distributions of strongly prognostic factors may vary between the trial populations and the treated patient population. Thus, if the EBCTCG estimate is interpreted as the marginal hazard ratio, differences in prognostic factor distributions provide reason to interpret the results with caution.

Due to uncertainty in the transportability of the hazard ratio from the EBCTCG to the treated patient population, an illustrative sensitivity analysis was conducted where the assumed target-population marginal hazard ratio was varied by $\pm 0.05$ around the EBCTCG estimate. These results are presented in Figure S2 and Figure S3. In these settings, the estimated overall ATT ranged from -9.2\% to -5.9\%, demonstrating that while the direction of the estimated ATT stayed the same, the magnitude is sensitive to departures from the marginal hazard ratio transportability assumption. 

\section{Discussion}\label{Discussion}
Randomised trials are typically the gold standard for estimating treatment effects, but the composition of patients in trials presents challenges for the external validity of trial results. Due to this limitation, generalisability and transportability methods have been established to make inference to populations whose composition differs from the trial population. Such methods have been extended to the meta-analysis setting when individual patient data are available. Here, the setting is considered where the only available information from a single trial or meta-analysis is the estimated relative treatment effect (i.e., risk ratio or hazard ratio), perhaps stratified by one or more baseline covariates. This setting is common when data restrictions preclude access to individual participant trial data. We provide a set of identification conditions and estimators for the ATT under marginal and conditional transportability assumptions. When their respective assumptions hold, the proposed estimators and variance estimation approach provide valid estimates of the ATT in the treated patient population and appropriately propagate error both from the trial(s) and from the patient cohort. 

The simulation study demonstrated that the conditions leading to violations of the transportability assumptions differed for risk ratios and hazard ratios. For risk ratios, a collapsible measure, the transportability assumption held when treatment effect heterogeneity was appropriately accounted for in estimation. This was true for the data generating mechanisms considered here regardless of the strength of prognostic factors that differed in distribution between the trial population(s) and the target population. For the hazard ratio, a noncollapsible measure, the transportability assumption could be violated either by inappropriately accounting for treatment effect heterogeneity or by the presence of strong prognostic factors that differed in distribution between populations. These findings shed light on the settings where these transportability assumptions may be implausible and the proposed estimators may be biased. A recent meta-epidemiological study found that relative treatment effects are often not transportable across populations with varying baseline risks \citep{murad_variability_2024}. For these reasons, strong subject-matter knowledge and knowledge surrounding the data sources is required to make these transportability assumptions. In general, caution is required when transporting hazard ratios, as there is no clear rule of thumb regarding the degree of population differences or strength of prognostic factors required to produce meaningful bias. Similar concerns regarding transportability of hazard ratios have been raised previously \citep{didelez2022logic}.

The proposed methods are intended primarily for settings where individual participant data from the trials are unavailable such that transportability relies on trial summary measures. When reliable individual participant data and covariates  from trials are available, existing transportability approaches will generally be more flexible than those proposed here because they can account for a larger set of covariates than is typically available in trial summaries. The quality and precision of the trial summaries are also important. Simulations for the single trial setting demonstrate considerable finite sample bias when trial estimates were imprecise, with bias diminishing as trial sample sizes increased. Furthermore, uncertainty in the trial findings must be accounted for when estimating the variance of the ATT in the target population. Failure to account for this source of variation led to considerable confidence interval undercoverage in simulations. 

Beyond noncollapsibility, hazard ratios have other well-known limitations, particularly around interpretability \citep{hernan_hazards_2010}. Here, we provide identifiability conditions under which hazard ratios can be transformed into a more interpretable estimand in the target population. However, prior to application the validity of the transportability assumption must be carefully examined. The methods for transporting the hazard ratio to a target population also rely on the proportional hazards assumption. Such a strong assumption may not hold in many applications. Future work could consider the setting of transporting hazard ratios when the proportional hazards assumption is violated. Because trial follow-up is typically shorter than follow-up available in routinely collected data, hazard ratios should not be extrapolated beyond the length of follow-up supported by the trial data.   

The methods presented here were applied to estimate the ATT of endocrine therapy on cumulative risk of breast cancer death at 15 years among patients diagnosed in England between 2000-2009 aged 20-50 at the time of diagnosis. For the resulting estimates to be valid, the hazard ratio reported by the EBCTCG meta-analysis must appropriately represent the trial-population marginal hazard ratio, and this marginal hazard ratio must be transportable to the treated patient population.  Three considerations are particularly relevant to the plausibility of this assumption. First, the EBCTCG estimate was obtained using stratified log-rank analyses rather than as an unadjusted marginal hazard ratio, whereas estimator (\ref{EST.marg.HR}) specifies a marginal trial-population hazard ratio.  Second, this assumption would be violated if there is heterogeneity in relative treatment effects for measured or unmeasured patient or tumour-level characteristics that differ in distribution between the trial populations and the treated patient population. The EBCTCG meta-analysis found little evidence that relative effects of endocrine therapy varied by patient and tumour characteristics \citep{early_breast_cancer_trialists_collaborative_group_ebctcg_relevance_2011}, but the assumption of no heterogeneity is nonetheless strong. Third, even in the absence of treatment effect heterogeneity, differences in the distribution of strong prognostic factors between the trial populations and the treated patient population could result in different marginal hazard ratios due to noncollapsibility. The latter concern is particularly relevant as both the number of positive nodes and chemotherapy treatment, strong prognostic factors, may differ in distribution across populations. These features correspond to the setting in simulations in which noncollapsibility led to bias in estimates of the ATT. Consequently, there is uncertainty in treating the EBCTCG summary hazard ratio as the marginal target-population hazard ratio. Sensitivity analyses demonstrated the magnitude of differences in the ATT across moderate departures from the marginal transportability assumption.

There are other considerations for interpretation of the application. The cancer registry data are subject to incomplete recording of both oestrogen-receptor status and cancer treatments. For this reason, women were included in the treated patient population who were confirmed to be oestrogen-receptor positive or had recorded endocrine therapy, except those known to be oestrogen-receptor negative or with recorded use of aromatase inhibitors, a different type of endocrine therapy than tamoxifen. This inclusion criterion is unlikely to result in considerable bias because tamoxifen was the first-line treatment for oestrogen-receptor positive breast cancer among premenopausal women during this time period and was not recommended for oestrogen-receptor negative breast cancer patients. Similar measurement error concerns are common with real-world patient data and should be considered when applying these methods. This analysis also assumes that the level of compliance in the patient population is similar to the observed compliance in the tamoxifen trials, i.e., approximately 80\% \citep{early_breast_cancer_trialists_collaborative_group_ebctcg_relevance_2011}. Endocrine therapy prescription data provide some support for the plausibility of this assumption \citep{Emanuel2019Endocrine}. However, patient adherence to a five-year regimen is difficult to measure in both routinely-collected data and randomised trials, and non-adherence has been associated with poorer patient outcomes \citep{Eliassen2023Importance}. Despite its limitations, this analysis demonstrates the proposed methods and shows variation in absolute treatment effects when the hazard ratio is assumed to be marginally transportable to the treated patient population. 

There are several potential extensions to this work. The estimators presented here assume non-informative censoring in the patient population. These estimators could be easily extended to accommodate inverse-probability of censoring weights for the setting where there is censoring by observed covariates in the general patient population. Because of the ubiquity of competing events in many epidemiological settings, including the application in Section 4, future work could consider identification assumptions for transportability when competing events are present. 

When used appropriately, this method provides an approach for combining summary measures from randomised trials with representative data from a target population receiving treatment for estimation of absolute treatment effects. Such absolute treatment effect estimates of the benefits and harms of different treatment options are beneficial for clinicians and patients when making informed treatment decisions. While motivated by estimation of the absolute effect of endocrine therapy for breast cancer, these methods can be applied in other settings when individual participant trial data are unavailable and the identifiability conditions are reasonably satisfied.

\section*{Supplementary Material} 
R code for computing the different estimators and for the simulation study described in Section 3 is available at \url{https://github.com/bonnieshook/Transporting_Trial_Summary_Measures}. 

\section*{Bibliography}

	\bibliographystyle{apalike}
	\bibliography{bibliography}

\begin{thebibliography}{}

\bibitem[Andersen, 2021]{andersen_absolute_2021}
Andersen, L.~W. (2021).
\newblock Absolute vs. relative effects—implications for subgroup analyses.
\newblock {\em Trials}, 22(1):50.

\bibitem[Balduzzi et~al., 2019]{balduzzi_how_2019}
Balduzzi, S., Rücker, G., and Schwarzer, G. (2019).
\newblock How to perform a meta-analysis with {R}: a practical tutorial.
\newblock {\em Evidence-Based Mental Health}, 22(4):153--160.

\bibitem[Breskin et~al., 2019]{breskin_using_2019}
Breskin, A., Westreich, D., Cole, S.~R., and Edwards, J.~K. (2019).
\newblock Using {Bounds} to {Compare} the {Strength} of {Exchangeability} {Assumptions} for {Internal} and {External} {Validity}.
\newblock {\em American Journal of Epidemiology}, 188(7):1355--1360.

\bibitem[Campbell and Jansen, 2026]{campbell2026hidden}
Campbell, H. and Jansen, J.~P. (2026).
\newblock Hidden in plain sight: How non-collapsibility biases treatment effects in (network) meta-analysis.
\newblock {\em arXiv:2603.00749}.

\bibitem[Dahabreh et~al., 2020]{dahabreh_toward_2020}
Dahabreh, I.~J., Petito, L.~C., Robertson, S.~E., Hernán, M.~A., and Steingrimsson, J.~A. (2020).
\newblock Toward {Causally} {Interpretable} {Meta}-analysis: {Transporting} {Inferences} from {Multiple} {Randomized} {Trials} to a {New} {Target} {Population}.
\newblock {\em Epidemiology}, 31(3):334.

\bibitem[Dahabreh et~al., 2023]{dahabreh_efficient_2023}
Dahabreh, I.~J., Robertson, S.~E., Petito, L.~C., Hernán, M.~A., and Steingrimsson, J.~A. (2023).
\newblock Efficient and {Robust} {Methods} for {Causally} {Interpretable} {Meta}-{Analysis}: {Transporting} {Inferences} from {Multiple} {Randomized} {Trials} to a {Target} {Population}.
\newblock {\em Biometrics}, 79(2):1057--1072.

\bibitem[Dahabreh et~al., 2024]{dahabreh_learning_2024}
Dahabreh, I.~J., Robertson, S.~E., and Steingrimsson, J.~A. (2024).
\newblock Learning about treatment effects in a new target population under transportability assumptions for relative effect measures.
\newblock {\em European Journal of Epidemiology}, 39(9):957--965.

\bibitem[Dahabreh et~al., 2019]{dahabreh_generalizing_2019}
Dahabreh, I.~J., Robertson, S.~E., Tchetgen, E.~J., Stuart, E.~A., and Hernán, M.~A. (2019).
\newblock Generalizing {Causal} {Inferences} from {Individuals} in {Randomized} {Trials} to {All} {Trial}-{Eligible} {Individuals}.
\newblock {\em Biometrics}, 75(2):685--694.

\bibitem[Daniel et~al., 2021]{daniel2021making}
Daniel, R., Zhang, J., and Farewell, D. (2021).
\newblock Making apples from oranges: Comparing noncollapsible effect estimators and their standard errors after adjustment for different covariate sets.
\newblock {\em Biometrical Journal}, 63(3):528--557.

\bibitem[Didelez and Stensrud, 2022]{didelez2022logic}
Didelez, V. and Stensrud, M.~J. (2022).
\newblock On the logic of collapsibility for causal effect measures.
\newblock {\em Biometrical Journal}, 64(2):235--242.

\bibitem[{Early Breast Cancer Trialists' Collaborative Group}, 2011]{early_breast_cancer_trialists_collaborative_group_ebctcg_relevance_2011}
{Early Breast Cancer Trialists' Collaborative Group} (2011).
\newblock Relevance of breast cancer hormone receptors and other factors to the efficacy of adjuvant tamoxifen: patient-level meta-analysis of randomised trials.
\newblock {\em The Lancet}, 378(9793):771--784.

\bibitem[Eliassen et~al., 2023]{Eliassen2023Importance}
Eliassen, F.~M., Bl{\aa}fjelldal, V., Helland, T., Hjorth, C.~F., H{\o}lland, K., Lode, L., Bertelsen, B.~E., Janssen, E.~A., Mellgren, G., Kval{\o}y, J.~T., and S{\o}iland, H. (2023).
\newblock Importance of endocrine treatment adherence and persistence in breast cancer survivorship: a systematic review.
\newblock {\em BMC Cancer}, 23(1):625.

\bibitem[Emanuel et~al., 2019]{Emanuel2019Endocrine}
Emanuel, G., Henson, K.~E., Broggio, J., Charman, J., Horgan, K., Dodwell, D., and Darby, S.~C. (2019).
\newblock Endocrine therapy in the years following a diagnosis of breast cancer: a proof of concept study using the primary care prescription database linked to cancer registration data.
\newblock {\em Cancer Epidemiology}, 61:185--189.

\bibitem[Furukawa et~al., 2002]{furukawa_can_2002}
Furukawa, T.~A., Guyatt, G.~H., and Griffith, L.~E. (2002).
\newblock Can we individualize the ‘number needed to treat’? {An} empirical study of summary effect measures in meta-analyses.
\newblock {\em International Journal of Epidemiology}, 31(1):72--76.

\bibitem[Hernán, 2010]{hernan_hazards_2010}
Hernán, M.~A. (2010).
\newblock The {Hazards} of {Hazard} {Ratios}.
\newblock {\em Epidemiology}, 21(1):13.

\bibitem[Jones et~al., 2025]{jones_inequalities_2025}
Jones, D.~A., Spencer, K., Ramroth, J., Probert, J., Roope, L. S.~J., Shakir, R., Broggio, J., Burroughs, F., Collins, G.~P., Clarke, P.~M., Wolstenholme, J.~L., and Cutter, D.~J. (2025).
\newblock Inequalities in geographic barriers and patient representation in lymphoma clinical trials across {England}.
\newblock {\em British Journal of Haematology}, 206(2):531--540.

\bibitem[Karanis et~al., 2016]{karanis_research_2016}
Karanis, Y.~B., Canta, F. A.~B., Mitrofan, L., Mistry, H., and Anger, C. (2016).
\newblock ‘{Research}’ vs ‘real world’ patients: the representativeness of clinical trial participants.
\newblock {\em Annals of Oncology}, 27:vi542.

\bibitem[Kent et~al., 2016]{Kent_IJE}
Kent, D.~M., Nelson, J., Dahabreh, I.~J., Rothwell, P.~M., Altman, D.~G., and Hayward, R.~A. (2016).
\newblock Risk and treatment effect heterogeneity: re-analysis of individual participant data from 32 large clinical trials.
\newblock {\em International Journal of Epidemiology}, 45(6):2075--2088.

\bibitem[Murad et~al., 2024]{murad_variability_2024}
Murad, M.~H., Wang, Z., Xiao, M., Chu, H., and Lin, L. (2024).
\newblock Variability of relative treatment effect among populations with low, moderate and high control group event rates: a meta-epidemiological study.
\newblock {\em BMC Medical Research Methodology}, 24(1):263.

\bibitem[Murad et~al., 2023]{murad_methods_2023}
Murad, M.~H., Wang, Z., Zhu, Y., Saadi, S., Chu, H., and Lin, L. (2023).
\newblock Methods for deriving risk difference (absolute risk reduction) from a meta-analysis.
\newblock {\em BMJ}, 381.

\bibitem[{National Cancer Institute}, 2026]{nci_hormone_therapy_breast_cancer}
{National Cancer Institute} (2026).
\newblock Hormone therapy | breast cancer treatment.
\newblock Accessed 30 June 2026.

\bibitem[Phillippo et~al., 2026]{phillippo2026multilevel}
Phillippo, D.~M., Dias, S., Ades, A., and Welton, N.~J. (2026).
\newblock Multilevel network meta-regression for general likelihoods: synthesis of individual and aggregate data with applications to survival analysis.
\newblock {\em Journal of the Royal Statistical Society Series A: Statistics in Society}, 189(3):1856--1875.

\bibitem[Remiro-Az{\'o}car, 2024]{remiro2024transportability}
Remiro-Az{\'o}car, A. (2024).
\newblock Transportability of model-based estimands in evidence synthesis.
\newblock {\em Statistics in Medicine}, 43(22):4217--4249.

\bibitem[Schmid et~al., 1998]{schmid_empirical_1998}
Schmid, C.~H., Lau, J., McIntosh, M.~W., and Cappelleri, J.~C. (1998).
\newblock An empirical study of the effect of the control rate as a predictor of treatment efficacy in meta-analysis of clinical trials.
\newblock {\em Statistics in Medicine}, 17(17):1923--1942.

\bibitem[Taylor et~al., 2023]{taylor_breast_2023}
Taylor, C., McGale, P., Probert, J., Broggio, J., Charman, J., Darby, S.~C., Kerr, A.~J., Whelan, T., Cutter, D.~J., Mannu, G., and Dodwell, D. (2023).
\newblock Breast cancer mortality in 500 000 women with early invasive breast cancer diagnosed in {England}, 1993-2015: population based observational cohort study.
\newblock {\em BMJ}, 381:e074684.

\bibitem[Westreich et~al., 2017]{westreich_transportability_2017}
Westreich, D., Edwards, J.~K., Lesko, C.~R., Stuart, E., and Cole, S.~R. (2017).
\newblock Transportability of {Trial} {Results} {Using} {Inverse} {Odds} of {Sampling} {Weights}.
\newblock {\em American Journal of Epidemiology}, 186(8):1010--1014.

\end{thebibliography}

\section*{Acknowledgements} 
Funding was provided by Cancer Research UK (PRCRPG-Nov21\textbackslash{}100001) and the University of Oxford. These funding bodies had no role in study design; in the collection, analysis, and interpretation of data; in the writing of the report; or in the decision to submit the article for publication.

A University of Oxford institutional version of ChatGPT 5.4 was used to suggest language revisions to improve clarity and readability, to identify relevant related work, to provide coding support, and to format Figure S1. The authors are solely responsible for the content. \vspace*{-8pt}

	\clearpage
	
	\begin{appendices}
		\setcounter{equation}{0}
		\renewcommand{\theequation}{S.\arabic{equation}}

    \setcounter{table}{0}
    \renewcommand{\thetable}{S\arabic{table}}

    \setcounter{figure}{0}
    \renewcommand{\thefigure}{S\arabic{figure}}

\section*{S1: Proof of identifiability of $ATT_{R=0}(t)$ under Assumption 6(a), i.e., marginal transportability of the risk ratio}

\label{sec:App.RR.marg}

Note that Assumption 6(a) implies that $\Pr(T(0) < t \mid A=1,R=0)=\Pr(T(1)< t \mid A=1,R=0) \{ RR_{R>0}(t) \}^{-1}$. Therefore, under the stated assumptions,
\begin{align}
ATT_{R=0}(t)
&= \Pr(T(1)<t \mid A=1,R=0) - \Pr(T(0)<t \mid A=1,R=0) \notag \\
&= \Pr(T(1)<t \mid A=1,R=0)
   - \Pr(T(1)<t \mid A=1,R=0) \{ RR_{R>0}(t) \}^{-1} \notag \\
&= \Pr(T<t \mid A=1,R=0)
   - \Pr(T<t \mid A=1,R=0) \{ RR_{R>0}(t) \}^{-1} \notag \\
&= \Pr(T<t \mid A=1,R=0)
   \left[1-\{ RR_{R>0}(t) \}^{-1} \right] \notag
\end{align}
where the first equality holds by definition of $ATT_{R=0}(t)$ and the second by Assumption 6(a). The third equality holds by Assumption 2, and the final equality holds by algebraic manipulation.

\section*{S2: Proof of identifiability of $ATT_{R=0}(t)$ under Assumption 6(b), i.e., conditional transportability of the risk ratio}

By algebraic manipulation, Assumption 6(b) implies that \[
\begin{aligned}
\Pr(T(0)<t \mid A=1, R=0, X=x)
&= \Pr(T(1)<t \mid A=1, R=0, X=x) \\
&\qquad {}\times \{RR_{R>0}(t \mid X=x)\}^{-1}
\end{aligned}
\] for all $x$ such that $\Pr(X=x \mid A=1, R=0)>0$. Then note that
\begin{align}
ATT_{R=0}(t)
&=
\Pr(T(1)<t \mid A=1, R=0)
-
\Pr(T(0)<t \mid A=1, R=0) \notag \\
&=
\sum_x
\Big[
\Pr(T(1)<t \mid A=1, R=0, X=x)
-
\Pr(T(0)<t \mid A=1, R=0, X=x)
\Big] \notag \\
&\qquad\qquad \times \Pr(X=x \mid A=1, R=0) \notag \\
&=
\sum_x
\Big[
\Pr(T(1)<t \mid A=1, R=0, X=x) \notag \\
&\qquad -
\Pr(T(1)<t \mid A=1, R=0, X=x)
\{RR_{R>0}(t \mid X=x)\}^{-1}
\Big] \notag \\
&\qquad\qquad \times \Pr(X=x \mid A=1, R=0) \notag \\
&=
\sum_x
\Big[
\Pr(T<t \mid A=1, R=0, X=x) \notag \\
&\qquad -
\Pr(T<t \mid A=1, R=0, X=x)
\{RR_{R>0}(t \mid X=x)\}^{-1}
\Big] \notag \\
&\qquad\qquad \times \Pr(X=x \mid A=1, R=0) \notag \\
&=
\sum_x
\Pr(T<t \mid A=1, R=0, X=x)
\left[
1-\{RR_{R>0}(t \mid X=x)\}^{-1}
\right] \notag \\
&\qquad\qquad \times \Pr(X=x \mid A=1, R=0) \notag
\label{proof.ID.cond.RR}
\end{align}

\noindent{The} first equality holds by the definition of $ATT_{R=0}(t)$, the second equality
holds by the law of total probability, and the third by Assumption 6(b).
The fourth equality holds by Assumption 2, and the final equality holds by algebraic
manipulation.

\section*{S3: Proof of identifiability of $ATT_{R=0}(t)$ under Assumption 6(c), i.e., marginal transportability of the hazard ratio}
\begin{align}
ATT_{R=0}(t)
&= \Pr(T(1)<t \mid A=1,R=0) - \Pr(T(0)<t \mid A=1,R=0) \notag \\
&= \{1-S_1(t \mid A=1,R=0) \}
   - \{1-S_0(t \mid A=1,R=0) \}  \notag \\
&= \{S_1(t \mid A=1,R=0) \}^{HR_{R=0,A=1}^{-1}}
   - S_1(t \mid A=1,R=0)  \notag \\
&= \{S_1(t \mid A=1,R=0) \}^{HR_{R>0}^{-1}}
   - S_1(t \mid A=1,R=0)  \notag \\
&= \{S(t \mid A=1,R=0) \}^{HR_{R>0}^{-1}}
   - S(t \mid A=1,R=0)  \notag
\end{align}

\noindent{where} $S(t \mid A=1,R=0)= \Pr(T \geq t \mid  A=1,R=0)$. Here, the first and second equalities hold by definition of $ATT_{R=0}(t)$ and $S_a(t \mid A=1,R=0)$, respectively, and the third by the proportional hazards assumption. The fourth equality holds by Assumption 6(c), and the final equality holds by Assumption 2.

\section*{S4: Proof of identifiability of $ATT_{R=0}(t)$ under Assumption 6(d), i.e., conditional transportability of the hazard ratio}
\begin{align}
ATT_{R=0}(t)
&= \Pr(T(1)<t \mid A=1, R=0)
   - \Pr(T(0)<t \mid A=1, R=0) \notag \\
&= \{1 - S_1(t \mid A=1, R=0)\}
   - \{1 - S_0(t \mid A=1, R=0)\} \notag \\
&= S_0(t \mid A=1, R=0)
   - S_1(t \mid A=1, R=0) \notag \\
&= \sum_x
   \left\{S_0(t \mid A=1, R=0, X=x)
   - S_1(t \mid A=1, R=0, X=x)\right\} \notag \\
&\qquad {}\times \Pr(X=x \mid A=1, R=0) \notag \\
&= \sum_x
   \left\{S_1(t \mid A=1, R=0, X=x)^{HR^{-1}_{R=0,X=x}}
   - S_1(t \mid A=1, R=0, X=x)\right\} \notag \\
&\qquad {}\times \Pr(X=x \mid A=1, R=0) \notag \\
&= \sum_x
   \left\{S_1(t \mid A=1, R=0, X=x)^{HR^{-1}_{R>0,X=x}}
   - S_1(t \mid A=1, R=0, X=x)\right\} \notag \\
&\qquad {}\times \Pr(X=x \mid A=1, R=0) \notag \\
&= \sum_x
   \left\{S(t \mid A=1, R=0, X=x)^{HR^{-1}_{R>0,X=x}}
   - S(t \mid A=1, R=0, X=x)\right\} \notag \\
&\qquad {}\times \Pr(X=x \mid A=1, R=0) \notag 
\end{align}

\noindent{where} the first two equalities hold by the definition of $ATT_{R=0}(t)$ and
$S_a(t \mid A=1, R=0)$, the third equality holds by algebraic manipulation,
the fourth by the law of total probability, and the fifth by the proportional
hazards assumption. The sixth equality holds by Assumption 6(d) and the final equality holds by Assumption 2.

\begin{table}
\begin{threeparttable}
\caption{Simulation summary results, transporting summary measures from a meta-analysis of 15 trials, naïve variance estimator, 2000 simulations}
\label{tab.S1}
\centering
\small
\setlength{\tabcolsep}{3.5pt} 

\begin{tabular}{l c l c c c c c c} 
\hline
  \makecell{Prognostic \\ Factors}  & Scenario & Estimator & \makecell{Bias \\ (x100)} & \makecell{Relative \\ Bias(\%)} & ASE & ESE & SER & \makecell{95\% CI \\ Coverage} \\ 
  \hline 
 Moderate & 1A & Marginal & 0.0 & 0.5 & 0.04 & 1.13 & 0.04 & 6 \\
 & 1A & Conditional & 0.0 & 1.8 & 0.05 & 1.47 & 0.03 & 4 \\
 & 1B & Marginal & 0.5 & -15.2 & 0.05 & 1.11 & 0.04 & 7 \\
 & 1B & Conditional & 0.0 & 1.0 & 0.06 & 1.49 & 0.04 & 5 \\
 & 1C & Marginal & 1.5 & -32.3 & 0.06 & 1.07 & 0.05 & 4 \\
 & 1C & Conditional & 0.0 & 1.0 & 0.10 & 1.55 & 0.06 & 10 \\
 & 1D & Marginal & 0.4 & -12.7 & 0.05 & 1.11 & 0.04 & 6 \\
 & 1D & Conditional & 0.4 & -11.6 & 0.05 & 1.45 & 0.04 & 6 \\
 & 1E & Marginal & 1.2 & -28.9 & 0.05 & 1.08 & 0.05 & 5 \\
 & 1E & Conditional & 1.1 & -28.0 & 0.06 & 1.42 & 0.04 & 6 \\
   \hline
Strong & 1A & Marginal & 0.0 & 0.8 & 0.05 & 1.47 & 0.03 & 5 \\
 & 1A & Conditional & -0.1 & 2.2 & 0.05 & 1.71 & 0.03 & 5 \\
 & 1B & Marginal & 0.7 & -15.6 & 0.05 & 1.45 & 0.04 & 6 \\
 & 1B & Conditional & 0.0 & 1.0 & 0.07 & 1.75 & 0.04 & 6 \\
 & 1C & Marginal & 2.2 & -32.0 & 0.07 & 1.40 & 0.05 & 3 \\
 & 1C & Conditional & -0.1 & 1.0 & 0.12 & 1.86 & 0.06 & 10 \\
 & 1D & Marginal & 0.6 & -14.1 & 0.05 & 1.44 & 0.04 & 5 \\
 & 1D & Conditional & 0.6 & -13.0 & 0.06 & 1.68 & 0.03 & 5 \\
 & 1E & Marginal & 1.9 & -31.2 & 0.06 & 1.39 & 0.04 & 3 \\
 & 1E & Conditional & 1.8 & -30.3 & 0.06 & 1.63 & 0.04 & 3 \\
  \hline
Moderate & 2A & Marginal & 0.0 & 0.5 & 0.08 & 1.26 & 0.07 & 11 \\
 & 2A & Conditional & 0.1 & -0.8 & 0.08 & 1.54 & 0.05 & 10 \\
 & 2B & Marginal & 1.1 & -9.5 & 0.09 & 1.25 & 0.07 & 8 \\
 & 2B & Conditional & 0.1 & -0.9 & 0.09 & 1.57 & 0.06 & 10 \\
 & 2C & Marginal & 3.6 & -30.3 & 0.07 & 1.18 & 0.06 & 0 \\
 & 2C & Conditional & 0.1 & -1.2 & 0.11 & 1.60 & 0.07 & 12 \\
 & 2D & Marginal & 0.8 & -7.4 & 0.08 & 1.25 & 0.07 & 8 \\
 & 2D & Conditional & 1.0 & -8.5 & 0.09 & 1.53 & 0.06 & 6 \\
 & 2E & Marginal & 2.6 & -25.7 & 0.06 & 1.19 & 0.05 & 1 \\
 & 2E & Conditional & 2.7 & -26.4 & 0.06 & 1.49 & 0.04 & 2 \\
  \hline
Strong & 2A & Marginal & -1.3 & 12.2 & 0.03 & 1.30 & 0.02 & 2 \\
 & 2A & Conditional & -0.6 & 5.6 & 0.07 & 1.28 & 0.05 & 7 \\
 & 2B & Marginal & -0.3 & 2.7 & 0.04 & 1.30 & 0.03 & 4 \\
 & 2B & Conditional & -0.7 & 5.7 & 0.07 & 1.29 & 0.06 & 8 \\
 & 2C & Marginal & 2.1 & -17.9 & 0.03 & 1.28 & 0.03 & 1 \\
 & 2C & Conditional & -0.7 & 6.1 & 0.08 & 1.30 & 0.06 & 7 \\
 & 2D & Marginal & -0.6 & 5.4 & 0.03 & 1.29 & 0.03 & 4 \\
 & 2D & Conditional & 0.2 & -1.4 & 0.07 & 1.29 & 0.05 & 8 \\
 & 2E & Marginal & 1.2 & -12.4 & 0.03 & 1.26 & 0.02 & 3 \\
 & 2E & Conditional & 1.9 & -19.1 & 0.05 & 1.27 & 0.04 & 3 \\
 \hline
\end{tabular}
\begin{tablenotes}
      \small
      \item Note: Bias, relative bias, ASE, ESE, SER, and 95\% CI coverage are calculated for the ATT. ASE, SER, and 95\% CI coverage are based on the naïve approach for estimating the variance, where variation in meta-analyses are not taken into account. 
Abbreviations: ASE = Average standard error; ESE = empirical standard error; SER = standard error ratio; CI = confidence interval

    \end{tablenotes}
    \end{threeparttable}
\end{table}

\begin{table} 
\begin{threeparttable}
\caption{Simulation summary results, transporting summary measures from a single trial of 650 participants, 2000 simulations}
\label{tab:ST600}
\centering
\small
\setlength{\tabcolsep}{3.5pt} 

\begin{tabular}{l c l c c c c c c} 
\hline
  \makecell{Prognostic \\ Factors}  & Scenario & Estimator & \makecell{Bias \\ (x100)} & \makecell{Relative \\ Bias(\%)} & ASE & ESE & SER & \makecell{95\% CI \\ Coverage} \\ 
  \hline 
 Moderate & 1A & Marginal & -0.6 & 23.9 & 4.58 & 4.61 & 0.99 & 95 \\
 & 1A & Conditional & -1.3 & 50.0 & 5.66 & 5.60 & 1.01 & 97 \\
 & 1B & Marginal & -0.2 & 6.4 & 4.54 & 4.58 & 0.99 & 95 \\
 & 1B & Conditional & -1.3 & 39.9 & 5.81 & 5.78 & 1.00 & 96 \\
 & 1C & Marginal & 0.6 & -12.3 & 4.45 & 4.49 & 0.99 & 92 \\
 & 1C & Conditional & -1.5 & 31.6 & 6.20 & 6.10 & 1.02 & 96 \\
 & 1D & Marginal & -0.2 & 7.1 & 4.52 & 4.56 & 0.99 & 94 \\
 & 1D & Conditional & -0.9 & 28.9 & 5.60 & 5.55 & 1.01 & 96 \\
 & 1E & Marginal & 0.6 & -14.8 & 4.39 & 4.44 & 0.99 & 92 \\
 & 1E & Conditional & -0.1 & 1.5 & 5.46 & 5.39 & 1.01 & 94 \\
  \hline 
 Strong & 1A & Marginal & -0.6 & 17.0 & 5.76 & 5.85 & 0.98 & 95 \\
 & 1A & Conditional & -1.0 & 28.4 & 6.28 & 6.26 & 1.00 & 96 \\
 & 1B & Marginal & 0.0 & 0.4 & 5.72 & 5.83 & 0.98 & 94 \\
 & 1B & Conditional & -1.0 & 21.6 & 6.44 & 6.47 & 1.00 & 96 \\
 & 1C & Marginal & 1.1 & -15.6 & 5.62 & 5.73 & 0.98 & 92 \\
 & 1C & Conditional & -1.2 & 17.5 & 6.86 & 6.87 & 1.00 & 96 \\
 & 1D & Marginal & 0.1 & -1.3 & 5.68 & 5.81 & 0.98 & 94 \\
 & 1D & Conditional & -0.3 & 7.8 & 6.19 & 6.20 & 1.00 & 95 \\
 & 1E & Marginal & 1.3 & -22.1 & 5.50 & 5.63 & 0.98 & 90 \\
 & 1E & Conditional & 0.9 & -15.5 & 6.02 & 6.01 & 1.00 & 92 \\
\hline
Moderate & 2A & Marginal & -0.5 & 4.5 & 4.74 & 4.87 & 0.97 & 95 \\
 & 2A & Conditional & -0.7 & 6.4 & 5.08 & 5.35 & 0.95 & 94 \\
 & 2B & Marginal & 0.4 & -3.7 & 4.71 & 4.85 & 0.97 & 93 \\
 & 2B & Conditional & -0.7 & 6.0 & 5.16 & 5.45 & 0.95 & 94 \\
 & 2C & Marginal & 2.4 & -20.3 & 4.52 & 4.66 & 0.97 & 89 \\
 & 2C & Conditional & -0.7 & 6.2 & 5.22 & 5.57 & 0.94 & 93 \\
 & 2D & Marginal & 0.4 & -3.7 & 4.69 & 4.85 & 0.97 & 93 \\
 & 2D & Conditional & 0.2 & -1.8 & 5.04 & 5.35 & 0.94 & 93 \\
 & 2E & Marginal & 2.3 & -22.2 & 4.48 & 4.62 & 0.97 & 90 \\
 & 2E & Conditional & 2.1 & -20.1 & 4.86 & 5.15 & 0.94 & 90 \\
 \hline
Strong & 2A & Marginal & -1.4 & 13.5 & 4.77 & 4.93 & 0.97 & 93 \\
 & 2A & Conditional & -0.8 & 7.8 & 4.33 & 4.58 & 0.95 & 93 \\
 & 2B & Marginal & -0.8 & 6.8 & 4.79 & 4.95 & 0.97 & 94 \\
 & 2B & Conditional & -0.9 & 8.0 & 4.35 & 4.59 & 0.95 & 92 \\
 & 2C & Marginal & 0.9 & -7.9 & 4.72 & 4.89 & 0.96 & 94 \\
 & 2C & Conditional & -1.0 & 8.4 & 4.33 & 4.58 & 0.94 & 92 \\
 & 2D & Marginal & -0.7 & 6.4 & 4.76 & 4.93 & 0.97 & 94 \\
 & 2D & Conditional & -0.1 & 0.7 & 4.32 & 4.57 & 0.94 & 93 \\
 & 2E & Marginal & 1.1 & -11.4 & 4.65 & 4.83 & 0.96 & 93 \\
 & 2E & Conditional & 1.7 & -17.2 & 4.24 & 4.51 & 0.94 & 92 \\

\end{tabular}
\begin{tablenotes}
      \small
      \item Note: Bias, relative bias, ASE, ESE, SER, and 95\% CI coverage are calculated for the ATT at $t=15$. ASE, SER, and 95\% CI coverage are based on the Monte-Carlo procedure with 500 replications. 
Abbreviations: ASE = Average standard error; ESE = empirical standard error; SER = standard error ratio; CI = confidence interval

    \end{tablenotes}
    \end{threeparttable}
\end{table}

\begin{table} 
\begin{threeparttable}
\caption{Simulation summary results, transporting summary measures from a single trial of 1500 participants, 2000 simulations}
\label{tab:ST1500}
\centering
\small
\setlength{\tabcolsep}{3.5pt} 

\begin{tabular}{l c l c c c c c c} 
\hline
  \makecell{Prognostic \\ Factors}  & Scenario & Estimator & \makecell{Bias \\ (x100)} & \makecell{Relative \\ Bias(\%)} & ASE & ESE & SER & \makecell{95\% CI \\ Coverage} \\ 
  \hline 
 Moderate & 1A & Marginal & -0.1 & 4.1 & 2.90 & 2.89 & 1.00 & 95 \\
 & 1A & Conditional & -0.3 & 11.0 & 3.31 & 3.29 & 1.01 & 95 \\
 & 1B & Marginal & 0.3 & -9.0 & 2.87 & 2.86 & 1.00 & 94 \\
 & 1B & Conditional & -0.3 & 8.2 & 3.38 & 3.35 & 1.01 & 95 \\
 & 1C & Marginal & 1.1 & -22.7 & 2.81 & 2.80 & 1.00 & 90 \\
 & 1C & Conditional & -0.4 & 7.8 & 3.55 & 3.52 & 1.01 & 95 \\
 & 1D & Marginal & 0.3 & -9.6 & 2.86 & 2.86 & 1.00 & 93 \\
 & 1D & Conditional & 0.1 & -4.0 & 3.27 & 3.25 & 1.01 & 94 \\
 & 1E & Marginal & 1.1 & -26.8 & 2.78 & 2.77 & 1.00 & 90 \\
 & 1E & Conditional & 0.9 & -22.5 & 3.19 & 3.16 & 1.01 & 91 \\
   \hline 
 Strong & 1A & Marginal & -0.2 & 4.6 & 3.68 & 3.66 & 1.01 & 95 \\
 & 1A & Conditional & -0.4 & 10.0 & 3.88 & 3.88 & 1.00 & 96 \\
 & 1B & Marginal & 0.4 & -8.7 & 3.65 & 3.63 & 1.01 & 94 \\
 & 1B & Conditional & -0.3 & 7.2 & 3.96 & 3.98 & 1.00 & 96 \\
 & 1C & Marginal & 1.5 & -21.9 & 3.59 & 3.57 & 1.00 & 90 \\
 & 1C & Conditional & -0.4 & 6.0 & 4.17 & 4.15 & 1.01 & 96 \\
 & 1D & Marginal & 0.5 & -11.1 & 3.63 & 3.60 & 1.01 & 93 \\
 & 1D & Conditional & 0.3 & -6.8 & 3.82 & 3.82 & 1.00 & 94 \\
 & 1E & Marginal & 1.8 & -29.3 & 3.51 & 3.48 & 1.01 & 89 \\
 & 1E & Conditional & 1.6 & -26.3 & 3.71 & 3.69 & 1.01 & 90 \\
\hline
Moderate & 2A & Marginal & -0.2 & 1.7 & 3.12 & 3.14 & 0.99 & 95 \\
 & 2A & Conditional & -0.2 & 2.2 & 3.39 & 3.39 & 1.00 & 94 \\
 & 2B & Marginal & 0.7 & -6.1 & 3.10 & 3.12 & 0.99 & 94 \\
 & 2B & Conditional & -0.3 & 2.3 & 3.44 & 3.44 & 1.00 & 94 \\
 & 2C & Marginal & 2.6 & -22.5 & 2.98 & 3.01 & 0.99 & 84 \\
 & 2C & Conditional & -0.3 & 2.5 & 3.50 & 3.48 & 1.01 & 95 \\
 & 2D & Marginal & 0.7 & -6.3 & 3.08 & 3.11 & 0.99 & 93 \\
 & 2D & Conditional & 0.7 & -5.7 & 3.36 & 3.36 & 1.00 & 94 \\
 & 2E & Marginal & 2.5 & -24.5 & 2.95 & 2.98 & 0.99 & 85 \\
 & 2E & Conditional & 2.4 & -23.8 & 3.23 & 3.22 & 1.00 & 86 \\
 \hline
 Strong & 2A & Marginal & -1.3 & 12.0 & 3.16 & 3.17 & 1.00 & 93 \\
 & 2A & Conditional & -0.7 & 6.8 & 2.89 & 2.89 & 1.00 & 94 \\
 & 2B & Marginal & -0.6 & 5.2 & 3.17 & 3.18 & 1.00 & 94 \\
 & 2B & Conditional & -0.8 & 6.9 & 2.91 & 2.90 & 1.00 & 94 \\
 & 2C & Marginal & 1.1 & -9.4 & 3.12 & 3.17 & 0.99 & 92 \\
 & 2C & Conditional & -0.9 & 7.5 & 2.90 & 2.92 & 0.99 & 93 \\
 & 2D & Marginal & -0.6 & 4.9 & 3.16 & 3.18 & 0.99 & 94 \\
 & 2D & Conditional & 0.0 & -0.3 & 2.89 & 2.88 & 1.00 & 95 \\
 & 2E & Marginal & 1.3 & -12.8 & 3.08 & 3.14 & 0.98 & 92 \\
 & 2E & Conditional & 1.8 & -18.0 & 2.83 & 2.85 & 0.99 & 89 \\

\end{tabular}
\begin{tablenotes}
      \small
      \item Note: Bias, relative bias, ASE, ESE, SER, and 95\% CI coverage are calculated for the ATT at $t=15$. ASE, SER, and 95\% CI coverage are based on the Monte-Carlo procedure with 500 replications. 
Abbreviations: ASE = Average standard error; ESE = empirical standard error; SER = standard error ratio; CI = confidence interval

    \end{tablenotes}
    \end{threeparttable}
\end{table}

\begin{table} 
\begin{threeparttable}
\caption{Simulation summary results, transporting summary measures from a single trial of 5000 participants, 2000 simulations}
\label{tab:ST5000}
\centering
\small
\setlength{\tabcolsep}{3.5pt} 

\begin{tabular}{l c l c c c c c c} 
\hline
  \makecell{Prognostic \\ Factors}  & Scenario & Estimator & \makecell{Bias \\ (x100)} & \makecell{Relative \\ Bias(\%)} & ASE & ESE & SER & \makecell{95\% CI \\ Coverage} \\ 
  \hline 
 Moderate & 1A & Marginal & 0.0 & 0.5 & 1.57 & 1.58 & 0.99 & 95 \\
 & 1A & Conditional & -0.1 & 2.8 & 1.75 & 1.75 & 1.00 & 95 \\
 & 1B & Marginal & 0.4 & -11.5 & 1.55 & 1.57 & 0.99 & 93 \\
 & 1B & Conditional & -0.1 & 1.8 & 1.78 & 1.79 & 0.99 & 95 \\
 & 1C & Marginal & 1.2 & -24.7 & 1.52 & 1.54 & 0.99 & 85 \\
 & 1C & Conditional & -0.1 & 1.7 & 1.85 & 1.87 & 0.99 & 95 \\
 & 1D & Marginal & 0.4 & -12.6 & 1.55 & 1.56 & 0.99 & 93 \\
 & 1D & Conditional & 0.3 & -10.8 & 1.72 & 1.72 & 1.00 & 93 \\
 & 1E & Marginal & 1.2 & -28.7 & 1.51 & 1.51 & 0.99 & 85 \\
 & 1E & Conditional & 1.1 & -27.4 & 1.68 & 1.67 & 1.00 & 87 \\
 \hline
 Strong & 1A & Marginal & 0.0 & 0.9 & 2.00 & 2.00 & 1.00 & 95 \\
 & 1A & Conditional & -0.1 & 2.7 & 2.07 & 2.05 & 1.01 & 95 \\
 & 1B & Marginal & 0.5 & -11.6 & 1.98 & 1.98 & 1.00 & 93 \\
 & 1B & Conditional & -0.1 & 1.2 & 2.11 & 2.10 & 1.01 & 95 \\
 & 1C & Marginal & 1.7 & -23.9 & 1.95 & 1.94 & 1.00 & 83 \\
 & 1C & Conditional & -0.1 & 1.2 & 2.21 & 2.20 & 1.01 & 95 \\
 & 1D & Marginal & 0.6 & -13.9 & 1.97 & 1.97 & 1.00 & 93 \\
 & 1D & Conditional & 0.6 & -12.6 & 2.04 & 2.02 & 1.01 & 94 \\
 & 1E & Marginal & 1.9 & -31.0 & 1.91 & 1.91 & 1.00 & 80 \\
 & 1E & Conditional & 1.8 & -29.9 & 1.98 & 1.96 & 1.01 & 81 \\
\hline
Moderate & 2A & Marginal & -0.1 & 0.9 & 1.71 & 1.75 & 0.98 & 95 \\
 & 2A & Conditional & 0.0 & 0.4 & 1.87 & 1.93 & 0.97 & 95 \\
 & 2B & Marginal & 0.8 & -6.7 & 1.70 & 1.74 & 0.98 & 90 \\
 & 2B & Conditional & -0.1 & 0.7 & 1.90 & 1.97 & 0.97 & 95 \\
 & 2C & Marginal & 2.7 & -23.2 & 1.63 & 1.68 & 0.97 & 61 \\
 & 2C & Conditional & -0.1 & 0.7 & 1.93 & 2.01 & 0.96 & 94 \\
 & 2D & Marginal & 0.8 & -7.0 & 1.69 & 1.73 & 0.98 & 90 \\
 & 2D & Conditional & 0.8 & -7.4 & 1.85 & 1.92 & 0.97 & 91 \\
 & 2E & Marginal & 2.6 & -25.3 & 1.62 & 1.66 & 0.98 & 63 \\
 & 2E & Conditional & 2.6 & -25.6 & 1.78 & 1.83 & 0.97 & 67 \\
 \hline
Strong & 2A & Marginal & -1.2 & 11.7 & 1.74 & 1.77 & 0.99 & 88 \\
 & 2A & Conditional & -0.7 & 6.6 & 1.60 & 1.62 & 0.99 & 92 \\
 & 2B & Marginal & -0.6 & 5.0 & 1.75 & 1.78 & 0.98 & 94 \\
 & 2B & Conditional & -0.8 & 6.8 & 1.61 & 1.64 & 0.98 & 92 \\
 & 2C & Marginal & 1.2 & -9.8 & 1.72 & 1.76 & 0.98 & 88 \\
 & 2C & Conditional & -0.9 & 7.3 & 1.60 & 1.65 & 0.98 & 91 \\
 & 2D & Marginal & -0.5 & 4.7 & 1.74 & 1.76 & 0.99 & 94 \\
 & 2D & Conditional & 0.1 & -0.5 & 1.60 & 1.61 & 0.99 & 95 \\
 & 2E & Marginal & 1.3 & -13.2 & 1.70 & 1.72 & 0.99 & 86 \\
 & 2E & Conditional & 1.8 & -18.4 & 1.56 & 1.58 & 0.99 & 77 \\
\hline
\end{tabular}
\begin{tablenotes}
      \small
      \item Note: Bias, relative bias, ASE, ESE, SER, and 95\% CI coverage are calculated for the ATT at $t=15$. ASE, SER, and 95\% CI coverage are based on the Monte-Carlo procedure with 500 replications. 
Abbreviations: ASE = Average standard error; ESE = empirical standard error; SER = standard error ratio; CI = confidence interval

    \end{tablenotes}
    \end{threeparttable}
\end{table}

\begin{figure}
\centering
\includegraphics[width=\columnwidth]{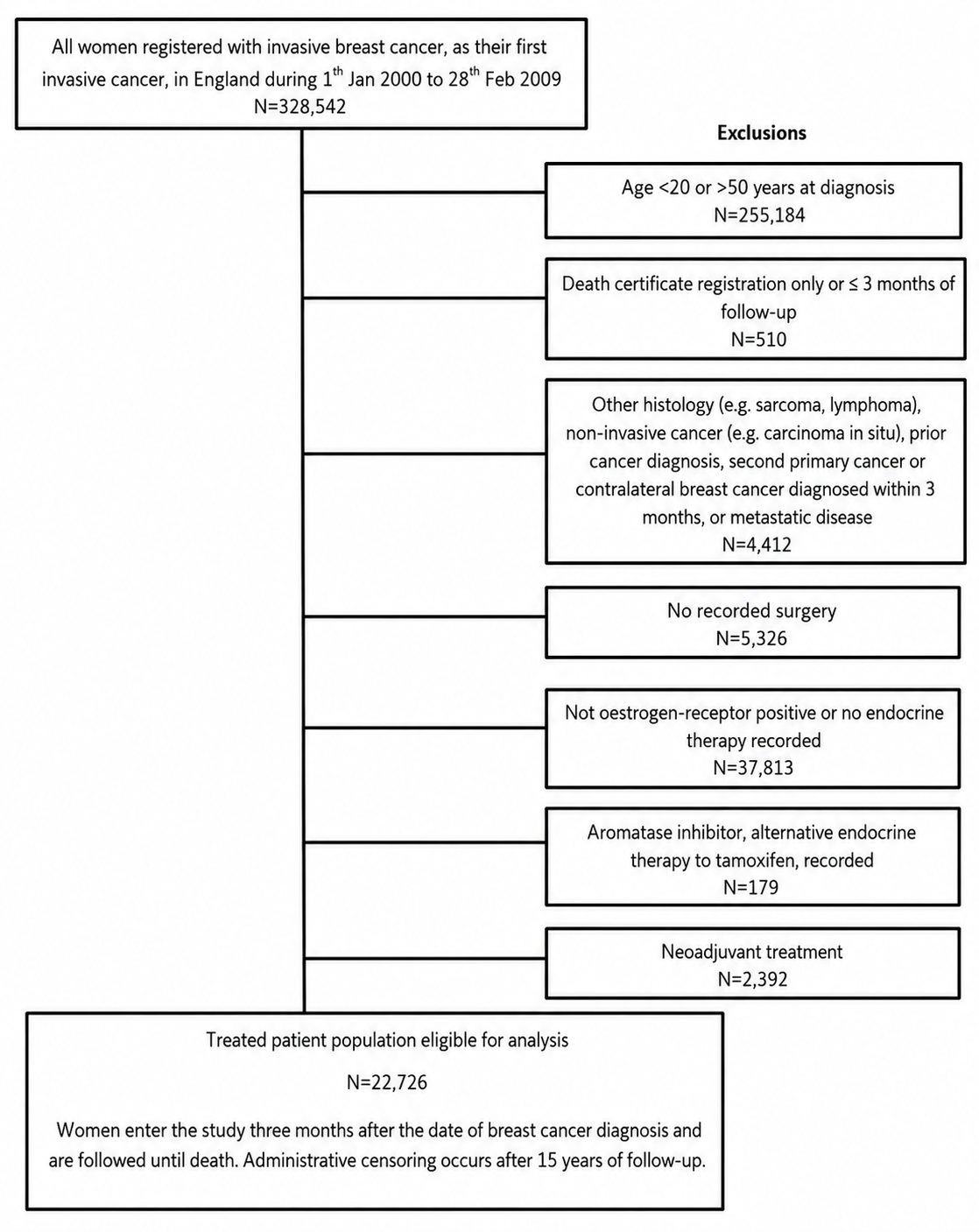}
\caption{Composition of the cohort}
\label{fig.Supp}
\end{figure}

\begin{table} 
\begin{threeparttable}
\caption{Unadjusted associational hazard ratios for breast cancer mortality for prognostic factors among women diagnosed with breast cancer in England during 2000-2009 at ages 20-50 years}
\label{tab:AppHRs}
\centering
\small
\setlength{\tabcolsep}{3.5pt} 

\begin{tabular}{l l c c c} 
\hline
  Characteristic &  & n & Events & \makecell{Hazard Ratio \\ 95\% CI} \\ 
  \hline 
Age at diagnosis (years)	&	20-44	&	10601	&	2259	&	1.00 (ref)	\\
	&	45-50	&	12125	&	1853	&	0.69 (0.65, 0.74)	\\
Calendar period of diagnosis	&	2000-2004	&	11883	&	2393	&	1.00 (ref)	\\
	&	2005-2009	&	10843	&	1719	&	0.76 (0.72, 0.81)	\\
Index of Multiple Deprivation	&	$<$20\% - least deprived	&	5201	&	831	&	1.00 (ref)	\\
	&	20-39\%	&	5084	&	848	&	1.05 (0.96, 1.16)	\\
	&	40-59\%	&	4732	&	860	&	1.16 (1.05, 1.27)	\\
	&	60-79\%	&	4257	&	856	&	1.30 (1.18, 1.43)	\\
	&	80\%$+$ - most deprived	&	3452	&	717	&	1.36 (1.23, 1.50)	\\
Geographical region	&	Eastern	&	3959	&	644	&	1.00 (ref)	\\
	&	North West	&	1048	&	200	&	1.19 (1.02, 1.40)	\\
	&	Northern and Yorkshire	&	5351	&	908	&	1.05 (0.95, 1.16)	\\
	&	Oxford	&	442	&	78	&	1.10 (0.87, 1.39)	\\
	&	South West	&	1074	&	193	&	1.11 (0.94, 1.30)	\\
	&	Thames	&	5285	&	1089	&	1.30 (1.18, 1.44)	\\
	&	Trent	&	1130	&	211	&	1.16 (1.00, 1.36)	\\
	&	West Midlands	&	4437	&	789	&	1.11 (1.00, 1.23)	\\
Ethnicity	&	White	&	20295	&	3595	&	1.00 (ref)	\\
	&	Asian or Asian British	&	692	&	128	&	1.05 (0.88, 1.25)	\\
	&	Black or Black British	&	516	&	135	&	1.58 (1.33, 1.88)	\\
	&	Other Ethnic Groups	&	330	&	51	&	0.85 (0.65, 1.12)	\\
	&	Mixed	&	110	&	20	&	1.04 (0.67, 1.61)	\\
	&	Unknown	&	783	&	183	&	1.51 (1.30, 1.75)	\\
Charlson comorbidity index	&	0	&	22182	&	3995	&	1.00 (ref)	\\
	&	1	&	448	&	92	&	1.18 (0.96, 1.45)	\\
	&	2$+$	&	96	&	25	&	1.71 (1.15, 2.53)	\\
Tumour size	&	1-20 mm	&	10989	&	1130	&	1.00 (ref)	\\
	&	21-50 mm	&	7499	&	1869	&	2.66 (2.47, 2.86)	\\
	&	$>$50 mm	&	935	&	374	&	4.86 (4.32, 5.46)	\\
	&	Unknown	&	3303	&	739	&	2.38 (2.17, 2.61)	\\
Number of positive nodes	&	0	&	5599	&	488	&	1.00 (ref)	\\
	&	1 to 3	&	4546	&	981	&	2.66 (2.39, 2.97)	\\
	&	4 to 9	&	1562	&	601	&	5.43 (4.82, 6.12)	\\
	&	10 or more	&	608	&	359	&	10.60 (9.25, 12.16)	\\
	&	Unknown	&	10411	&	1683	&	1.95 (1.77, 2.16)	\\
Chemotherapy	&	Not recorded	&	9198	&	891	&	1.00 (ref)	\\
	&	Recorded	&	13528	&	3221	&	2.69 (2.50, 2.90)	\\

\hline
\end{tabular}
\begin{tablenotes}
      \small
      \item Crude hazard ratios estimated from a univariable Cox proportional hazards model. 

Abbreviations: ref = reference level; CI = confidence interval

    \end{tablenotes}
    \end{threeparttable}
\end{table}

\begin{figure}[htbp]
\centering
\includegraphics[
  width=0.98\columnwidth,
  height=0.85\textheight,
  keepaspectratio,
  trim=4cm 18cm 4cm 18cm,
  clip
]{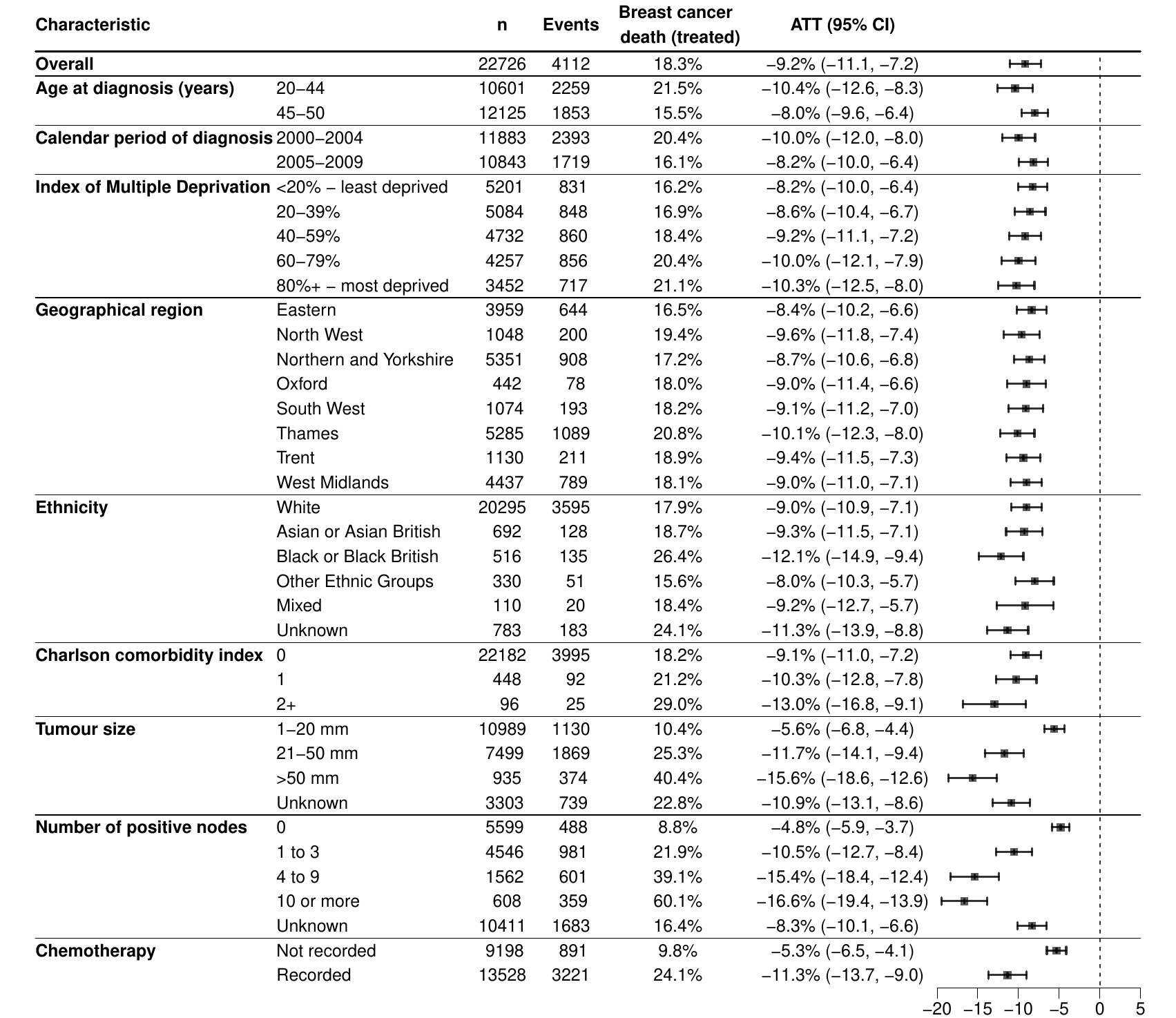}
\caption{Sensitivity Analysis 1: Estimated average treatment effect on the treated (ATT) of endocrine therapy on cumulative risk of breast cancer death at 15 years, overall and across patient subgroups, among women diagnosed with early breast cancer in England during 2000–2009 at ages 20–50 years assuming a marginal hazard ratio of 0.63. Note: Breast cancer death (treated) is the estimated 15-year cumulative risk of breast cancer death among patients receiving endocrine therapy. Negative ATT values indicate a lower 15-year cumulative risk of breast cancer death with endocrine therapy compared with no endocrine therapy.}
\end{figure}

\begin{figure}[htbp]
\centering
\includegraphics[
  width=0.98\columnwidth,
  height=0.85\textheight,
  keepaspectratio,
  trim=4cm 18cm 4cm 18cm,
  clip
]{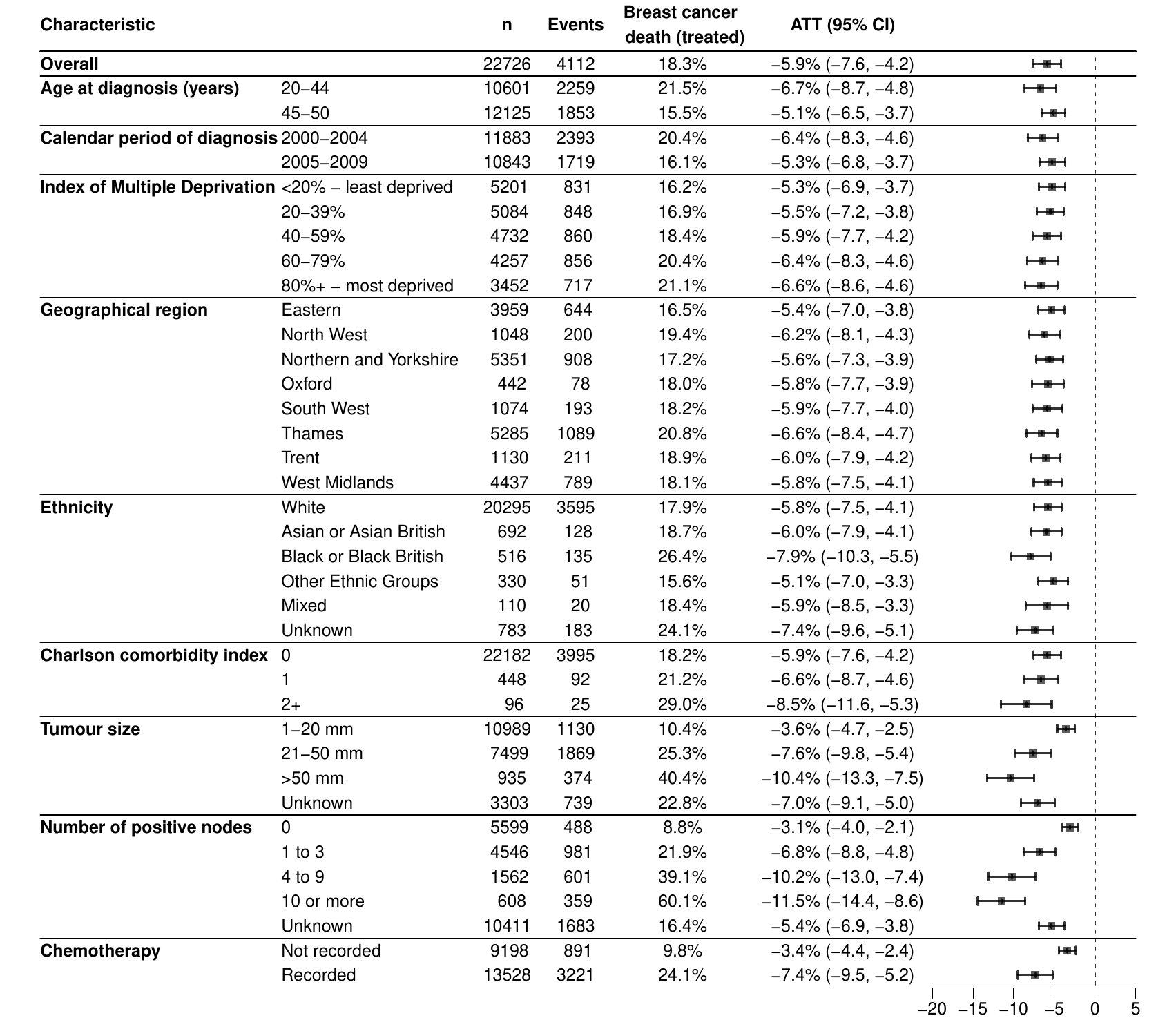}
\caption{Sensitivity Analysis 2: Estimated average treatment effect on the treated (ATT) of endocrine therapy on cumulative risk of breast cancer death at 15 years, overall and across patient subgroups, among women diagnosed with early breast cancer in England during 2000–2009 at ages 20–50 years assuming a marginal hazard ratio of 0.73. Note: Breast cancer death (treated) is the estimated 15-year cumulative risk of breast cancer death among patients receiving endocrine therapy. Negative ATT values indicate a lower 15-year cumulative risk of breast cancer death with endocrine therapy compared with no endocrine therapy.}
\end{figure}

	\end{appendices}	
	
	\label{lastpage}
	
\end{document}